\documentclass[conference]{IEEEtran}
\ifCLASSINFOpdf
\else
\fi

\usepackage{graphicx}
\usepackage{subcaption}
\usepackage{subfig} 
\usepackage{algorithm}
\usepackage{algorithmic}
\usepackage{url}
\usepackage{array}
\usepackage{cite}

\begin{document}
%
\title{ALPINE: Anytime Mining with Definite Guarantees}

\author{\IEEEauthorblockN{Qiong Hu\IEEEauthorrefmark{1} and
Tomasz Imielinski\IEEEauthorrefmark{1}}
\IEEEauthorblockA{\IEEEauthorrefmark{1}Department of Computer Science\\
Rutgers, The State University of New Jersey,
Piscataway, NJ 08854-8019\\Email: \{qionghu.cs, imielins\}@rutgers.edu}}




\maketitle

\begin{abstract}
ALPINE is to our knowledge the first anytime algorithm to mine frequent itemsets and closed frequent itemsets. It guarantees that all itemsets with support exceeding the current checkpoint's support have been found before it proceeds further. Thus, it is very attractive for extremely long mining tasks with very high dimensional data (for example in genetics) because it can offer intermediate meaningful and complete results. This ANYTIME feature is the most important contribution of ALPINE, which is also fast but not necessarily the fastest algorithm around. Another critical advantage of ALPINE is that it does not require the apriori decided minimum support value.
\end{abstract}


%
\IEEEpeerreviewmaketitle

\section{Introduction}
\label{section:introduction}

Finding all frequent patterns from large databases is NP-hard for it's an exhaustive search problem~\cite{zaki98:theoretical}. Almost all existing data mining technologies reviewed in~\cite{borgelt12:fism} are limited by the high response time due to the tough and compute-intensive nature of the task. As pointed out by Dass~\cite{dass05:bdfs}, all of the algorithms produce outputs only at the completion (either run to completion or provide no useful results) and are not amenable to the real-time decision-making need. 

For instance, consider a gene expression data set, i.e., microarray data, produced in bioinformatics\footnote{\url{http://www.broadinstitute.org/ccle/home}}, which usually have a large number of columns (genes). It's meaningful to find all the co-regulated genes or gene groups for (1) cancer treatment; (2) drug sensitivity analysis. However, it often cannot run to completion in a reasonable time for large microarray data. Imagine a mining process might be stuck computing for hours, days, even weeks without any response, it greatly challenges the user's patience and absolutely unacceptable in mission critical applications. Can we at least draw some partial conclusions during such a mining process without waiting until completion, i.e., first generate the higher support/coverage combination of genes which might have higher influence? Could an algorithm guarantee that all itemsets with support exceeding the current checkpoint's support have been found before it proceeds further? 


ALPINE is such an algorithm. Therefore, it exhibits so-called ANYTIME feature. An \emph{anytime algorithm} uses well-defined quality measures to monitor the progress in problem-solving and is expected to improve the quality of the solution as the computational time increases~\cite{zilberstein96:anytime}. Anytime algorithms have been categorized into two types: interruptible and contract algorithms~\cite{arai07:anytimetopk}. An interruptible algorithm can be interrupted at any time. A contract algorithm, if interrupted at any point before the termination of the contract time, might not yield any useful results. From this definition, an anytime algorithm is able to return many possible intermediate partial approximate answers to any given input. Thus, it is useful for solving problems where the search space is large and the quality of the results can be compromised~\cite{kranen09:anytime}. Clearly, the anytime approach is particularly well suited for data mining and more generally for intelligent systems.

\begin{figure}[t]
\centering
\includegraphics[width=0.5\textwidth]{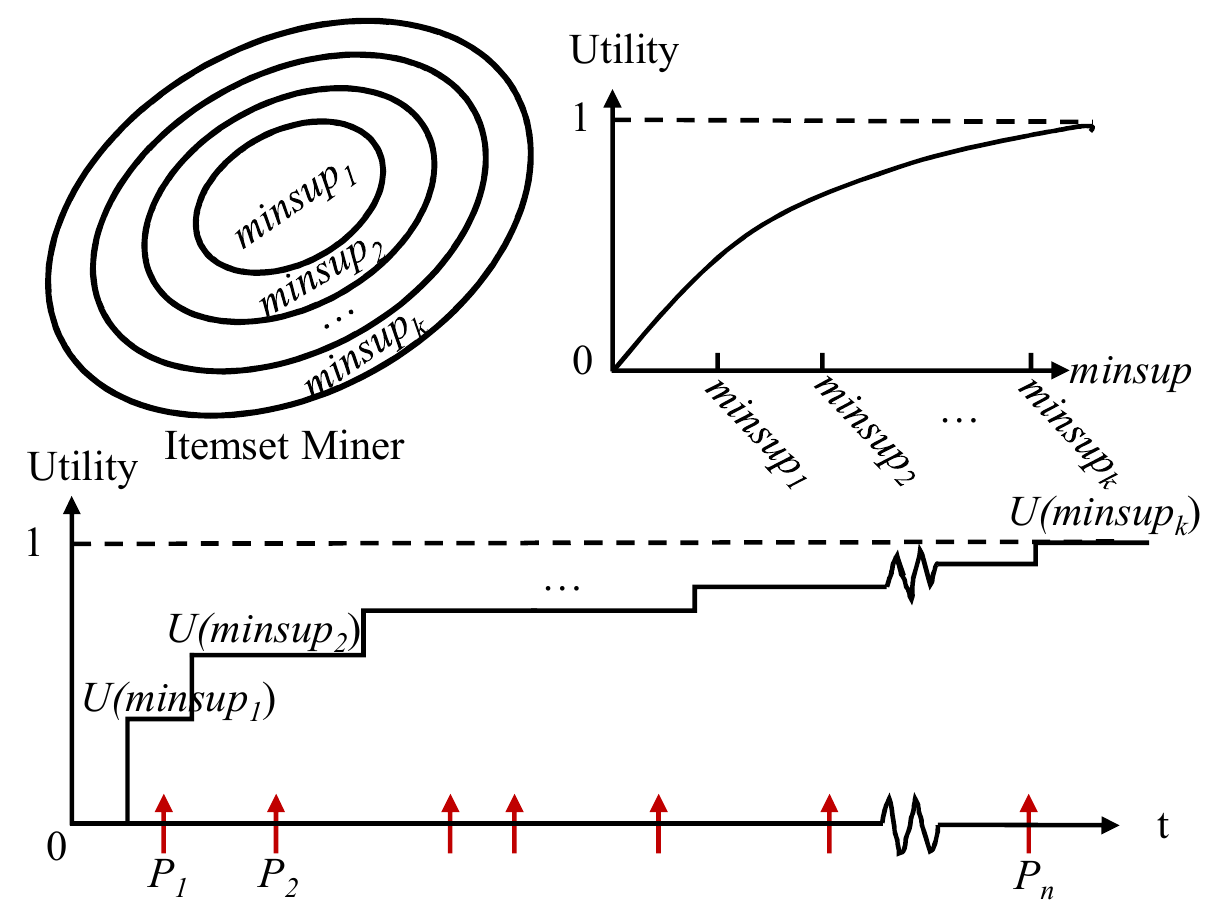}
\caption{The anytime mining framework. A long mining task, is progressively divided into $k$ sub-search spaces w.r.t. a set of decreasing minimum supports $\{minsup_{i = 1}^k\}$. The utility of the current solution is plotted as a function of time, where $\{P_{j = 1}^n\}$ is a set of randomly selected probes in time.
}
\label{fig:anytimeframework}
\end{figure}

In the context of frequent itemset mining, the common framework is to use a $minsup$ threshold to ensure the generation of the \emph{correct} and \emph{complete} set of patterns~\cite{wang05:tfp}. We require that an anytime mining algorithm reaches partial \emph{completeness} through checkpoints, which define the exploration of well-defined subspaces of the entire problem. According to the law of diminishing marginal utility in economics~\cite{rothbard56:law}, we believe that the additional benefit derived from the completeness of itemsets with a $minsup$ diminishes with the decrease of the value of $minsup$ (refer to the upper right graph in Figure~\ref{fig:anytimeframework}). The total utility derived from the outcome of an anytime mining algorithm can be utilized to quantify the usefulness of its intermediate results. The proposed anytime mining framework is illustrated in Figure 1, $\{minsup_{i = 1}^k\}$ is the set of all distinct supports of itemsets in a transaction database in decreasing order and $\{P_{j = 1}^n\}$ is a set of randomly selected probes in time. Upon the completion of all the itemsets with support greater than or equal to ${minsup_i}$, the utility associated with the minimum support, ${U(minsup_i)}$, is obtained instantly. The goal of an anytime miner is to maximize the average utility at the set of random probes, i.e., $\max \frac{1}{n} \sum_{j=1}^n{U(P_j)}$.

In this work, we are investigating anytime algorithms for frequent itemset mining and present the ALPINE algorithm, namely, Automatic $minsup$ Lowering with Progress Indicator in Never-Ending mining. The ALPINE algorithm proceeds in the defined anytime mining manner -- from checkpoint to checkpoint. In ALPINE, the checkpoints correspond to decreasing values of minimum support. ALPINE guarantees that all itemsets with support exceeding the current checkpoint's support have been found before it proceeds further. In this way, we know that we have completed a well-defined subset of the overall, potentially enormous search space. ALPINE proceeds in this \lq\lq monotonic\rq\rq~manner with minimal computational overhead as compared to the best existing frequent itemset mining algorithms. In ALPINE, though the mining process is continuous, it does not go totally unchecked.

ALPINE can be stopped at any point, we will always be able to offer partial conclusions based on the last checkpoint reached as indicated in Figure~\ref{fig:anytimeframework}. In contrast, the traditional itemset mining algorithms do not give any intermediate partial completeness guarantees, requiring the user to wait until completion to get any definite results.

There is another very critical advantage of ALPINE. It does not require setting the minimum support apriori. This requirement has always been problematic for all frequent itemset generation algorithms. How do we set the minimum support if we do not know the data? We only learn the data as we continue mining it, but then it is too late to change the value of minimum support. However, it is not the case for ALPINE.  ALPINE moves the minimum support as it goes ahead from checkpoint to checkpoint.  


ALPINE is, to our knowledge, the first anytime algorithm to mine frequent itemsets and closed frequent itemsets. Extensive experiments, with one of the fastest itemset mining algorithms in literature -- LCM~\cite{uno05:lcmv3}, illustrate the added value of this anytime feature and ALPINE\rq s minimal overhead compared with LCM. Since the sequential top-k mining algorithm, i.e., Seq-Miner~\cite{minh06:seq-miner}, enjoys some properties of the contract-type anytime algorithms, we also conduct a set of experiments to compare ALPINE with Seq-Miner.


\section{Preliminaries}
\label{section:preliminaries}

Let $\mathcal{I} = \{i_1, i_2, .., i_n\}$ be a set of literals, called \textit{items}, and $\mathcal{T} = \{t_1, t_2, ..., t_m\}$ be a \textit{transaction database}, where each transaction $t_k~( k = 1, 2, ..., m )$ in $\mathcal{T}$ is a set of items such that $t_k \subseteq \mathcal{I}$. A unique transaction identifier, $tid$, is associated with each transaction. Each subset of $\mathcal{I}$ is called an \textit{itemset} and a transaction is said to contain an itemset if all the items in the itemset are present in the transaction~\cite{agrawal94:apriori}. For an itemset $X$, the \textit{cover} $\mathcal{T}(X) = \{t \in \mathcal{T} | X \subseteq t\}$ (a $tidset$) be the set of transactions it is contained in and the support of $X$, denoted as $sup(X)$, is the number of these transactions. Hence,  $sup(X) = |\mathcal{T}(X)|$~\cite{borgelt12:fism}. If $sup(X)$ exceeds a minimum support threshold $minsup$, then $X$ is called a \textit{frequent itemset}~\cite{agrawal94:apriori, borgelt12:fism}. For a transaction set $\mathcal{S} \subseteq \mathcal{T}$, its intersection is $I(\mathcal{S}) = \cap_{T \in \mathcal{S}} T$. If an itemset $X$ satisfies $I(\mathcal{T}(X)) = X$, then $X$ is called a \textit{closed itemset}~\cite{pasquier99:closed}.

To avoid enumerating itemsets with duplications, it's natural to order the items to structure the search space. We define a total order among the set of items from $\mathcal{I}$ as item $i < $ item $j$ iff $sup(i) \leq sup(j)$.  The search is confined to extend an itemset only with items greater than all items inside it. Items from $\mathcal{I}$ can be recoded to 0, 1, ..., $|\mathcal{I}| - 1$ according to this order. Let $X = \{x_1, ..., x_n\}$ be an itemset as an ordered sequence such that $x_1 < ... < x_n$, the \textit{tail} of $X$ is $tail(X) = x_n$~\cite{uno03:lcm}. Then itemset $X$ will only be extended with all items greater than $tail(X)$,  resulting in a tree structured subset lattice.

To further reduce the search tree size, the closure operator $I(\mathcal{T}(\cdot))$ is utilized at each step. Together with the above extension rule, we define the \textit{closure} of itemset $X$ as $X^{*} = X \cup E$, where $E = \{e \in \mathcal{I} | e \in I(\mathcal{T}(X)) \land e > tail(X)\}$, for we know that $X$ union any subset of $E$ is supported exactly by $\mathcal{T}(X)$ ($E$ is a shortcut of support equivalence extension). Then $X$ is not needed to extend with items belong to $E$.  If $|E| = k$, this operation will reduce the size of the subtree rooted at $X$ by a factor of $2^k$. In general, for itemset $P$ and $Q$, with $P \subseteq Q \land \mathcal{T}(P) = \mathcal{T}(Q)$, the set of all itemset $Y$ which is a superset of $P$ and a subset of $Q$ can be compactly represented as an \textit{itemset interval}: $(P, Q) = \{Y | P \subseteq Y \subseteq Q\}$, for they all share the same supporting transactions as $P$. In this definition, $P$ and $Q$ specify the minimum itemset and the maximum itemset in an itemset interval, respectively.

We also define a mapping, $sup^{-1}$, from support to set of itemsets, which is applicable to both frequent and closed itemsets. We name it \emph{support index}, for it is used to index all the itemsets from a transaction database by support. Given a support value $s$, the indexed set of itemsets is $sup^{-1}(s) = \{X | sup(X) = s\}$. The degree of completeness for a specific support $s$, is defined as the number of itemsets discovered so far with support $s$ divided by the total number of itemsets with support $s$ from the transaction database, i.e., $|sup^{-1}(s)|$.

\section{The ALPINE Algorithm}
\label{section:alpine}

In ALPINE, itemsets are discovered in order of their supports -- from higher support to lower support. ALPINE will automatically lower the $minsup$ threshold to the next possible, lower value and continue mining. The basic idea of ALPINE is to dynamically build the support index, from the highest possible support gradually to the lowest possible value of support in the given transaction database. It progressively partitions itemset  intervals into disjoint bins of different supports. 

\begin{figure*}[t]
        \centering
        \begin{subfigure}[b]{0.46\textwidth}
                \includegraphics[width=\textwidth]{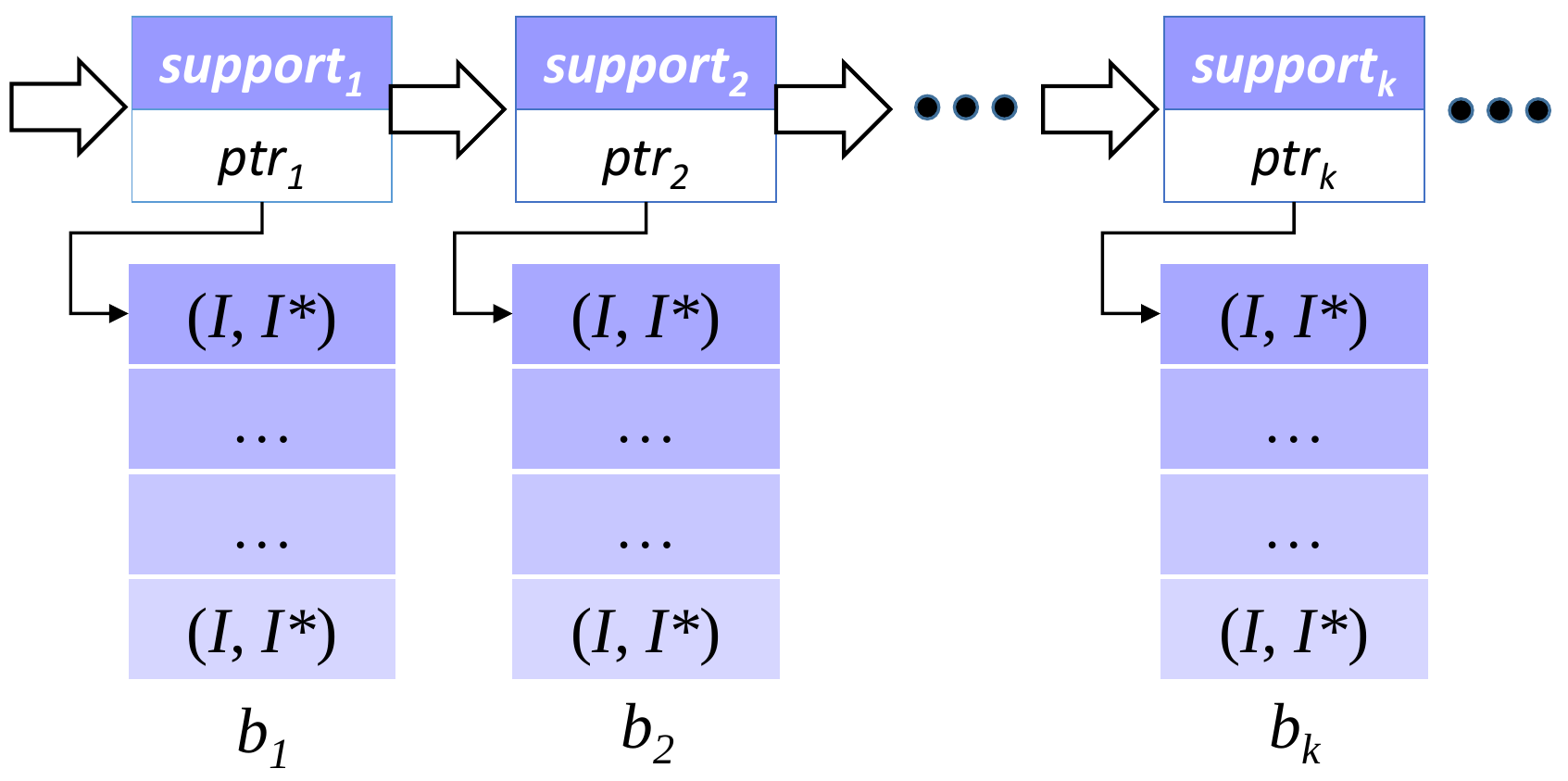}
                \caption{}
                \label{fig:frontier1}
        \end{subfigure}%
        \hskip .25cm 
        \begin{subfigure}[b]{0.48\textwidth}
                \includegraphics[width=\textwidth]{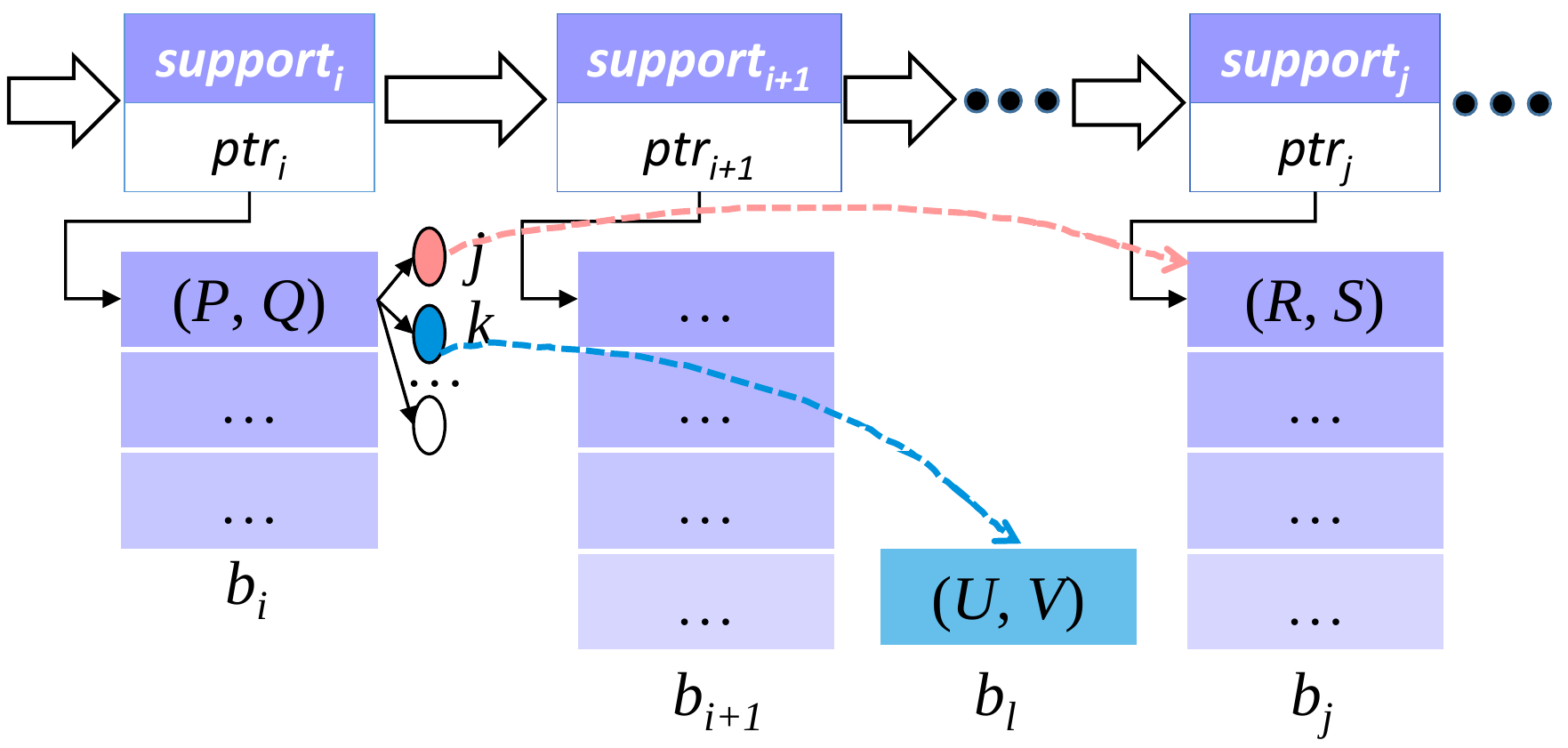}
                \caption{}
                \label{fig:frontier2}
        \end{subfigure}
        \caption{Dynamically index itemset intervals by their supports in the ALPINE algorithm}\label{fig:frontier}
\end{figure*}

ALPINE starts with the index built from all the itemset interval ($I$, $I^{*}$) of singleton itemset $I$ from a transaction database in Figure~\ref{fig:frontier1}. In this figure, all singleton itemset intervals are sorted in decreasing support from left to right and binned based on their support values. This index is not static, though, it is updated by new itemset intervals generated by extending the minimum itemset of an itemset interval. At any point in time, we are working on the uncompleted bin (there are still some remaining itemset intervals in this bin) with the highest support value $s$. To enumerate all itemsets with support above or equal to $s$, we extend the minimum itemset of each itemset interval in the bin. Let $(P, Q)$ be an itemset interval we are currently working on, we extend $P$ with all item $j$ greater than $tail(P)$ and not contained in $Q$. Denote $P \cup \{j\}$ as itemset $R$, we also find $R^{*}$ according to the definition of an itemset closure.  We can prove Lemma~\ref{lemma:interval} and get the itemset interval for $R$ accordingly: $(R, S)$, where $S = R^{*} \cup Q$ . 


\newtheorem{lemma}{Lemma}[section]
\begin{lemma}
All itemsets in the itemset interval $(R, S)$, where $S = R^{*} \cup Q$, are supported by $\mathcal{T}(R)$.
\label{lemma:interval}
\end{lemma}

\newtheorem{proof}{Proof}
\begin{proof}
$\forall$ itemset $X \in$ itemset interval $(R, S)$,  we can denote $X$ as $X = R \cup Y$, where $Y \subseteq S\setminus R$. For $S = R^{*} \cup Q$, we can further decompose $Y$ into two disjoint subsets $Y_R$ and $Y_Q$, with $Y_R \subseteq R^{*} \land Y_Q \subseteq Q \setminus R^{*}$. Then, the cover $\mathcal{T}(X) = \mathcal{T}(R \cup Y_R \cup Y_Q)$. According to the definition of $R^{*}$, we have $\mathcal{T}(R \cup Y_R) = \mathcal{T}(R)$, so $\mathcal{T}(X) = \mathcal{T}(R \cup Y_Q)$. Since $R = P \cup \{j\}$, then $\mathcal{T}(X) = \mathcal{T}(P \cup \{j\} \cup Y_Q)$. The itemset $P \cup Y_Q$ is in the itemset interval $(P, Q)$, so it is supported by $\mathcal{T}(P)$. Thus,  $\mathcal{T}(X) = \mathcal{T}(P \cup \{j\}) = \mathcal{T}(R)$.
\end{proof}


All of the newly generated itemset intervals $(R, S)$ are segregated by their support values into different bins of the support index. And $sup(R)$ will always be smaller than $sup(P)$ according to the closure operation, otherwise, item $j$ will belong to $Q$. There are two different situations: 1) the support of $R$  is already associated with  some existing bin and we only need to add itemset interval $(R, S)$ to that bin; 2) the support of $R$ is a new value, which hasn't been indexed yet and we need to create a new bin to place $(R, S)$ in it.

An example is given in Figure~\ref{fig:frontier2}, the itemset interval under exploration is on top of $b_i$ and $support_i$ corresponds to the highest uncompleted bin. Let's denote the interval as ($P, Q$). We extend $P$ with items greater than $tail(P)$ and not contained in $Q$. Suppose item $j$ and item $k$ belong to such set of items. When $P$ is extended with item $j$, we obtained  itemset $R = P \cup \{j\}$ and $S = R^{*}\cup Q$. It happens that $sup(R)$ is equal to some $support_j$ indexed, so the itemset interval ($R, S$) will be added to $b_j$ as indicated by the red dashed line in Figure~\ref{fig:frontier2}. However, when it comes to item $k$, the support of $U = P \cup \{k\}$ has not been indexed yet, a new bin $b_l$ is created to place its interval ($U, V$) with $V = U^{*} \cup Q$.

Only when we finish exploring all itemset intervals in a bin of the support index, we are safe to conclude that we have discovered all itemsets with support above or equal to the support associated with that bin. That's when we can declare that support as a new minimum support, $minsup$, and we have reached the successive checkpoint. This checkpoint completes the subspace of all itemsets with support above or equal to $minsup$, even though the bins with lower supports are not complete yet. ALPINE always continues to build new bins for the new possible support values or extends existing bins. In this way, ALPINE mines patterns with descending value of support sequentially and outputs partially complete information from checkpoint to checkpoint in the defined anytime mining manner. 

The pseudo-code of our prototypical algorithm \textrm{-} ALPINE is given in Algorithm~\ref{algorithm:ALPINE}. It starts from the bin of the highest support and it explores all itemset intervals in that bin one by one. Then it continues with the bins of successively lower support values. The init process in Line 1 initializes the support index from all the itemset intervals of the singleton itemsets in the given transaction database $\mathcal{T}$ in decreasing support order. Each itemset interval in a given bin is explored by the sub-routine Explorer given in Algorithm~\ref{algorithm:explorer}  (Line 5), in which the minimum itemset of each itemset interval is extended with all possible items in descending support order from the most promising one to the least promising one. Line 8 issues a checkpoint after completing all itemset intervals in a bin (with support above or equal to the support of that bin). The anytime feature makes the ALPINE algorithm can be interrupted at any moment (Line 9 - 11). From the described procedure, we can deduce the following lemma and observations:

\begin{algorithm}[htb]
\caption{ALPINE(Transactionset $\mathcal{T}$, Support index $\mathcal{S}$)} 
\label{algorithm:ALPINE} 
\begin{algorithmic}[1]
\STATE Init $\mathcal{S}$ by all itemset interval $(I, I^{*})$ of singleton itemset $I$ in $\mathcal{T}$;
\FORALL {$support_i$ of $\mathcal{S}$ in the decreasing order}
\STATE Get bin $b_i$ indexed by $support_i$;
\FORALL {itemset interval $(P, Q)$ in $b_i$}
\STATE Explorer($P$, $Q$, $\mathcal{S}$);
\ENDFOR
\STATE Declare $support_i$ to be $minsup$;
\STATE Issue checkpoint: complete subspace of all itemsets $\geq minsup$;
\IF {toTerminate == true}
\STATE break;   // terminate requested by user
\ENDIF
\ENDFOR
\end{algorithmic}
\end{algorithm}

\begin{algorithm}[htb]
\caption{Explorer(Minimum itemset $P$, Maximum itemset $Q$, Support index $\mathcal{S}$)}
\label{algorithm:explorer}
\begin{algorithmic}[1]
\STATE Output itemset interval: ($P, Q$);
\FORALL {item $j = |\mathcal{I}| - 1$; $j > tail(P)$; $j$-- --}
\IF {$j \in Q$}
\STATE continue; // no need to extend with item $\in Q$ 
\ENDIF
\STATE $R \leftarrow P\cup \{j\}$;
\STATE $S \leftarrow R^{*}\cup Q$;
\IF {$sup(R)$ is already indexed in $\mathcal{S}$}
\STATE Add $(R, S)$ to the indexed bin;
\ELSE
\STATE Create a new bin with support $sup(R)$ for $\mathcal{S}$ and add $(R, S)$ to it;
\ENDIF
\ENDFOR
\end{algorithmic}
\end{algorithm}

\medskip

\begin{lemma}
Each itemset will be output in exactly one itemset interval.
\end{lemma}

\begin{proof}
This lemma can be proved by contradiction. Suppose the same itemset $X$ can be output in two different itemset intervals: ($P_1, Q_1$) and ($P_2, Q_2$) with $P_1 \neq P_2$. Then we have, $P_1 \subseteq X \subseteq Q_1 \land P_2 \subseteq X \subseteq Q_2 \land \mathcal{T}(P_1) =  \mathcal{T}(P_2)$ (1). Thus, $P_1$ and $P_2$ cannot be inclusion relation with each other, otherwise, $\mathcal{T}(P_1) \neq \mathcal{T}(P_2)$. If we sort the items in $P_1$ and $P_2$ as ordered sequence according to the order defined in preliminaries and denote the first different item between them as $x_{1i}$ and $x_{2i}$ ($x_{1j} = x_{2j}, for~j = 0, 1, ..., i-1$, and we denote this set of items as $P_0$ and the associated itemset interval as ($P_0, Q_0$). Without loss of generality, we can assume $x_{1i} < x_{2i}$. From (1), we have $x_{1i} \in X \land x_{2i} \in X$. Consider itemset $R_1 = P_0 \cup \{x_{1i}\}$, since the itemset is only extended with items not in $Q_0$, we have $x_{1i} \notin Q_0$. Regarding itemset $R_2 = P_0 \cup \{x_{2i}\}$,  $S_2 = R_2^{*} \cup Q_0$. According to the definition of $R_2^{*}$, it doesn't include any item less than $x_{2i}$, so $x_{1i}$ must be in $Q_0$, which is a contradiction.
\end{proof}



\newtheorem{observation}{Observation}[section]

\begin{observation}
Every itemset with support above or equal to minsup was output at the related checkpoint.
\end{observation}


With the defined item order and the exploration procedure in Algorithm~\ref{algorithm:explorer}, each item greater than $tail(P)$ is either extended explicitly, or it is already contained in $Q$, which completes the search space. And all itemsets generated from the itemset intervals in the bins with lower supports are less than the $minsup$ of the current checkpoint according to the anti-monotone property of itemset support.


\begin{observation}
Every distinct support count of an itemset in the transaction database $\mathcal{T}$ will be $minsup$ value of some ALPINE's checkpoint.
\label{observation:checkpoint}
\end{observation}

This observation is readily obtained from the monotonic manner the ALPINE algorithm explores successive bins of the support index as discussed above. Thus,


\newtheorem{property}{Property}[section]
\begin{property}
Given a transaction database $\mathcal{T}$ with $m$ transactions over $n$ items, the number of checkpoints from it is bounded by $min\{2^n, m\}$.
\end{property}


\begin{property}
The minimum support of the first checkpoint is equal to the highest support of the singleton itemsets in $\mathcal{T}$.
\label{property:firstcheckpoint}
\end{property}

\vspace{-3mm}

\newtheorem{remark}{Remark}[section]
\begin{remark}
The highest support value which will correspond to the first and highest checkpoint for ALPINE is equal to the support of the most frequent item. If no other item shares that support, that item alone constitutes the first checkpoint. In this case, the subspace of itemsets corresponding to the first checkpoint has just one singleton set - that most frequent item.  
\end{remark}


\begin{remark}
It may be the case that several items share the same, highest support. In such case, ALPINE needs to do more work to reach the first checkpoint. In the extreme case, these top support items may be perfectly correlated (that is, all their combinations also have the same support). This is, of course, unlikely but possible. Suppose there are $k$ such items $i_{\pi_1}, i_{\pi_2}, ..., {i_{\pi_k}}$ sharing the highest value of support, and the order among them is $i_{\pi_1} < i_{\pi_2} < ... < i_{\pi_k}$. Then ALPINE needs to explore and output the following itemset intervals (by Algorithm~\ref{algorithm:explorer}) before reaching the first checkpoint: $(\{i_{\pi_1}\}, \{i_{\pi_1}i_{\pi_2}..i_{\pi_k}\})$, $(\{i_{\pi_2}\}, \{i_{\pi_2}i_{\pi_3}..i_{\pi_k}\})$, ..., $(\{i_{\pi_{k-1}}\}, \{i_{\pi_{k-1}}i_{\pi_k}\})$, $(\{i_{\pi_k}\}, \{i_{\pi_k}\})$, even though all the $k$ intervals are contained in ONE closed itemset $\{i_{\pi_1}i_{\pi_2}..i_{\pi_k}\}$.
\end{remark}


In this situation, it might be beneficial to confine the tree-shaped transversal routes of ALPINE to only closed itemsets, which can reduce the work for the aforementioned extreme case to explore only one itemset interval. In the next section, we'll show how to adapt ALPINE to mine closed itemsets.

\section{The ALPINEclosed Algorithm}
\label{section:alpineclosed}

The ALPINE algorithm elaborated in Section~\ref{section:alpine} can build the full support index of all itemsets from a transaction database $\mathcal{T}$ in decreasing support order. If we denote $\mathcal{F}$ and $\mathcal{C}$ the sets of all frequent itemsets and all frequent closed itemsets, respectively. According to their definitions, we know that $\mathcal{C}$ is a subset of $\mathcal{F}$. Thus, a straightforward way to adapt ALPINE for closed itemset mining is to add the closeness check for the maximum itemset of an itemset interval at each step of the mining process. 


\begin{observation}
Let $(P, Q)$ be an itemset interval, then $Q$ is closed if and only if $Q = I(\mathcal{T}(P))$.
\label{observation:closeness}
\end{observation}


From the definition of itemset interval, we have $Q$ is supported exactly by $\mathcal{T}(P)$. Thus, $Q = I(\mathcal{T}(P)) = I(\mathcal{T}(Q))$. According to the definition of closed itemset given in Section~\ref{section:preliminaries}, $Q$ is a closed itemset. 

For singleton itemset $I$, we know its itemset interval is $(I, I^{*})$. The closeness condition in Observation~\ref{observation:closeness} for $I^{*}$ is violated if and only if there exists some item $e < tail(I)$ such that $e$ occurs in every transaction in $\mathcal{T}(I)$. For an itemset interval $(R, S)$ generated from some itemset interval $(P, Q)$ in the intermediate stages of ALPINE, where $S = R^{*} \cup Q$, things are slightly different. To fail the closeness check, there must exist some item $e < tail(R) \land e \notin R$ satisfies: (1) $e$ is shared by every transaction of $\mathcal{T}(R)$; (2) $e \notin Q$. In other words, for closed itemset $S$, item $e < tail(R) \land e \notin R \land e \in I(\mathcal{T}(R))$ can only be obtained from $Q$. 

\begin{algorithm}[t]
\caption{ALPINEclosed(Transactionset $\mathcal{T}$, Support index $\mathcal{S}$)} 
\label{algorithm:ALPINEclosed} 
\begin{algorithmic}[1]
\STATE Init $\mathcal{S}$ by all itemset interval $(I, I^{*})$ of singleton itemset $I$ in $\mathcal{T}$ satisfying $I^{*} = I(\mathcal{T}(I))$ (closed);
\FORALL {$support_i$ of $\mathcal{S}$ in the decreasing order}
\STATE Get bin $b_i$ indexed by $support_i$;
\FORALL {itemset interval $(P, Q)$ in $b_i$}
\STATE Explorer2($P$, $Q$, $\mathcal{S}$);
\ENDFOR
\STATE Declare $support_i$ to be $minsup$;
\STATE Issue checkpoint: complete subspace of all \textit{closed} itemsets $\geq minsup$;
\IF {toTerminate == true}
\STATE break;   // terminate requested by user
\ENDIF
\ENDFOR
\end{algorithmic}
\end{algorithm}

\begin{algorithm}[t]
\caption{Explorer2(Minimum itemset $P$, Maximum itemset $Q$, Support index $\mathcal{S}$)}
\label{algorithm:explorer2}
\begin{algorithmic}[1]
\STATE Output closed itemset: $Q$;
\FORALL {item $j = |\mathcal{I}| - 1$; $j > tail(P)$; $j$-- --}
\IF {$j \in Q$}
\STATE continue; // no need to extend with item $\in Q$ 
\ENDIF
\STATE $R \leftarrow P\cup \{j\}$;
\STATE $S \leftarrow R^{*} \cup Q$;
\IF {$S = I(\mathcal{T}(R))$ (closed)}
\IF {$sup(R)$ is already indexed in $\mathcal{S}$}
\STATE Add $(R, S)$ to the indexed bin;
\ELSE
\STATE Create a new bin with support $sup(R)$ for $\mathcal{S}$ and add $(R, S)$ to it;
\ENDIF
\ENDIF
\ENDFOR
\end{algorithmic}
\end{algorithm}

With this observation, we can modify the ALPINE algorithm given in Section~\ref{section:alpine} to the ALPINEclosed algorithm (Algorithm~\ref{algorithm:ALPINEclosed}). Note that in Line 8 of Algorithm~\ref{algorithm:explorer2}, to test the closeness of itemset $S$, we only need to check all the item $k < tail(R) \land k \notin R \land k \notin Q$. For any nonempty closed itemset $S \in \mathcal{C}$, its parent is always defined and belongs to $\mathcal{C}$~\cite{uno03:lcm}. This guarantees the completeness of the proposed ALPINEclosed algorithm in mining all closed frequent itemsets. The difference of the ALPINE and ALPINEclosed algorithm lies in: 

\begin{itemize}
  \item Initialization: ALPINE is initialized with all itemset intervals of singleton itemsets from a database $\mathcal{T}$ (Line 1 of Algorithm~\ref{algorithm:ALPINE}). In contrast, ALPINEclosed is initialized with those itemset intervals passing the closeness test (Line 1 of Algorithm~\ref{algorithm:ALPINEclosed}).
  \item Exploration: ALPINE outputs itemset interval and keeps all newly generated itemset intervals $(R, S)$ in the mining process (Algorithm~\ref{algorithm:explorer}). ALPINEclosed, on the other hand, only outputs closed itemset and maintains itemset intervals $(R, S)$ meeting the closeness condition (Algorithm~\ref{algorithm:explorer2}).
  \item Status report: ALPINE reports the completion of the subspace of all frequent itemsets above the current $minsup$ (Line 8 of Algorithm~\ref{algorithm:ALPINE}). ALPINEclosed issues checkpoint about finishing all closed frequent itemsets above the current $minsup$ instead (Line 8 of Algorithm~\ref{algorithm:ALPINEclosed}).
\end{itemize}

\section{Computational Experiments}
In this section, we empirically evaluate the ALPINE algorithm and perform analysis in comparison with related works in both frequent itemset mining and sequential top-$k$ itemset mining. For frequent itemset generation, we choose one of the fastest itemset mining algorithms closely related with our work -- LCM (ver. 3)~\cite{uno05:lcmv3} and downloaded its implementation from the author's website\footnote{\url{http://research.nii.ac.jp/~uno/codes.htm}}. The utility gained at each probe of both algorithms can be used to quantify the usefulness of the intermediate partial solutions. As the measure of utility is usually application-dependent, we don't define the concrete utility function form here, but directly list the $minsup$ reached at each probe by both algorithms. In top-$k$ mining, we select the Seq-Miner~\cite{minh06:seq-miner} that mines the top-$k$ frequent patterns sequentially without any minimum support. The proposed ALPINE algorithm is implemented in JAVA and all the experiments were carried out on a cluster with 10 2.4 GHz processors and 256 GB memory. Both the experimental datasets from the FIMI repository\footnote{\url{http://fimi.ua.ac.be/data/}} and a real gene expression dataset from the Cancer Cell Line Encyclopedia (CCLE) project\footnote{\url{http://www.broadinstitute.org/ccle/data/browseData}} are used here.

\subsection{Comparison with frequent itemset mining}
\label{section:fim}

Let us start with emphasizing the benefits of anytime data mining. Since all conventional frequent itemset generation algorithms require setting up minimum support apriori by the user, what if the minimum support is set too low and the transaction database is too large? If this happens, such algorithms may run for very long time (practically forever), \lq\lq hanging\rq\rq~without providing any information to the user, except generating huge numbers of itemsets. However, no guarantees on the minimum support reached at each point are given. ALPINE, on the other hand, will provide the user with checkpoints which guarantee the partial completeness. It will provide lower and lower values of minimum support for which the set of frequent itemsets ALPINE produces is \emph{\textbf{complete}}. These guarantees will offer the user a measure of progress and knowledge about the subspace of the entire itemset search space that has been completely explored. A set of experiments is designed here to verify the added value of the anytime ALPINE algorithm and to check how ALPINE systematically explore the itemset space. We also analyze the computational overhead of ALPINE in time.


\subsubsection{Experiment on experimental datasets}
\label{section:fimi}

In the first set of experiments, we study the performance of ALPINE and LCM on experimental datasets from the FIMI repository. To illustrate the benefits of ALPINE, two relatively large transaction databases, i.e., T40I10D100K and Kosarak, which have many items and many transactions, are selected here. T40I10D100K has 100,000 transactions over 1,000 items generated by the IBM Quest Synthetic Data Generator, while Kosarak has 990,000 transactions over 41,270 items containing the click-stream data of a Hungarian on-line news portal. To reduce the number of mined itemsets, both ALPINE and LCM have confined to mine \emph{closed} itemsets  in this experiment. 

The experimental setting is as follows: we start both the ALPINE and the LCM algorithm at the same time, and probe every hour since they are started, i.e., Hour 1, Hour 2, ..., to check the status of both algorithms. Since ALPINE is parameter-free, it is not required to set any thresholds. It just continuously mines itemsets from checkpoint to checkpoint and tries to build the full support index for a given transaction database $\mathcal{T}$. Different from ALPINE, LCM must be initialized with some user-provided minimum support. In this experiment, we set the minimum support threshold of LCM to be 1 to mine all itemsets from $\mathcal{T}$ in consideration of building the full support index. The minimum support reached, i.e., all the itemsets with support greater than or equal to the minimum support are discovered, at each probe $t$ by the ALPINE algorithm is readily obtained from its last checkpoint before $t$, while this information for the LCM algorithm is obtained by post-processing all its output itemsets up to time $t$.

\begin{table}[b]
\caption{The $minsup$s reached at probes (in hour) by the LCM and ALPINE algorithm on the T40I10D100K (left) and Kosarak (right) dataset.}
\begin{tabular}{| m{.8cm} | m{.9cm} | m{1.2cm} | }
  \hline	
  \multicolumn{3}{|c|}{T40I10D100K } \\
  \hline
   Probe & LCM & ALPINE \\	
  \hline
  1  & 7314 & 116 \\
  2  & 6390 & 56 \\
  3  & 5855 & 34 \\
  4  & 5317 & 22 \\
  5  & 4873 & 16 \\ 
  6  & 4499 & 13 \\
  7  & 4168 & 11 \\
  8  & 3882 & 10 \\
  9  & 3575 & 9 \\
  10 & 3313 & 8 \\
  \hline  
\end{tabular}
~
\quad
\begin{tabular}{| m{.8cm} | m{.9cm} | m{1.2cm} | }
 \hline	
  \multicolumn{3}{|c|}{Kosarak } \\
  \hline
   Probe & LCM & ALPINE \\	
  \hline
   1 & 10178 & 982 \\
   2 & 9569 & 926 \\
   3 & 9264 & 907 \\
   4 & 8955 & 894 \\
   5 & 8810 & 885 \\
   6 & 8684 & 878 \\
   7 & 8645 & 872 \\
   8 & 8450 & 867 \\
   9 & 8379 & 862 \\
   10 & 8158 & 858 \\
   \hline
\end{tabular}
\label{table:fimi}
\end{table}


The results of both the algorithms for the set of probes up to ten hours on T40I10D100K and Kosarak datasets are shown in Table~\ref{table:fimi}. For both datasets, the first column is the probe time in hour, and the second and third column list the $minsup$ reached at each probe by LCM and ALPINE, respectively. It's clear from the table that ALPINE can quickly reach some lower minimum support value than LCM. For instance, on the T40I10D100K dataset, ALPINE can reach the minimum support of 116 in the first hour while LCM only completes the subspace of all itemsets with support above 7314. The same trend is also observed in the Kosarak dataset. For the Kosarak dataset has more items, it's even harder for the LCM algorithm to move minimum support. We notice that even after twenty hours, the minimum support LCM reached on the Kosarak dataset is 7920, while ALPINE has already finished all itemsets with support greater than or equal to 835.

The underlying reason is ALPINE systematically explore the itemset space in a \lq\lq monotonic\rq\rq~manner. ALPINE guarantees that all itemsets with support exceeding the current checkpoint's support have been found before it proceeds further, to build the support index for lower minimum support values. In contrast, LCM directly enumerate itemsets in a depth-first-search manner.  To understand how these two algorithms behavior differently, we have taken the partial solutions generated for the T40I10D100K dataset at one hour, three hours, six hours and ten hours as slices to look into the algorithms. We analyze these intermediate results and calculate the degree of completeness of all support levels. 

\begin{figure*}[t]
        \centering
        \begin{subfigure}[b]{0.49\textwidth}
                \includegraphics[width=\textwidth]{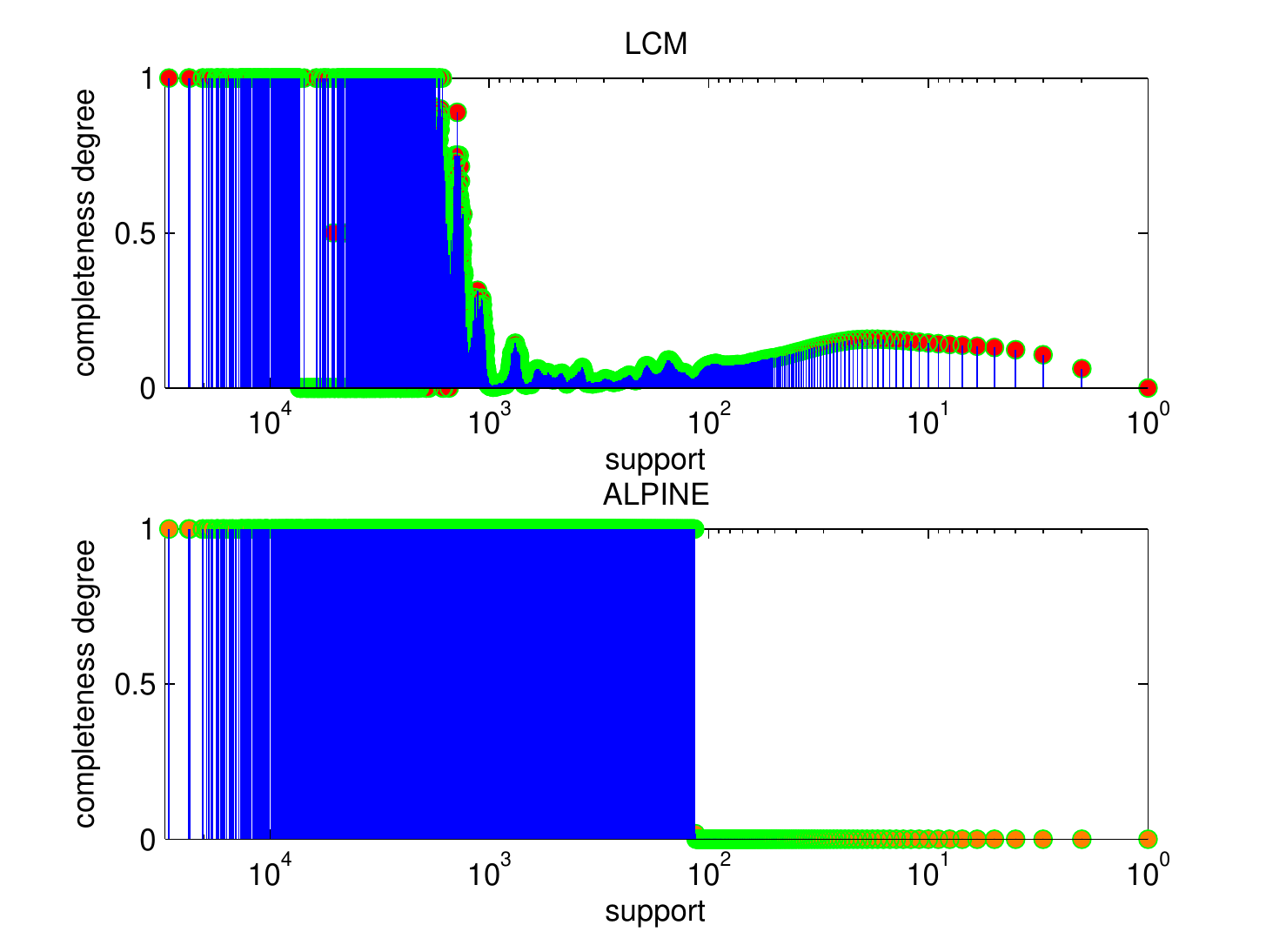}
                \caption{Hour 1}
                \label{fig:t40i10d100k1}
        \end{subfigure}%
        \begin{subfigure}[b]{0.49\textwidth}
                \includegraphics[width=\textwidth]{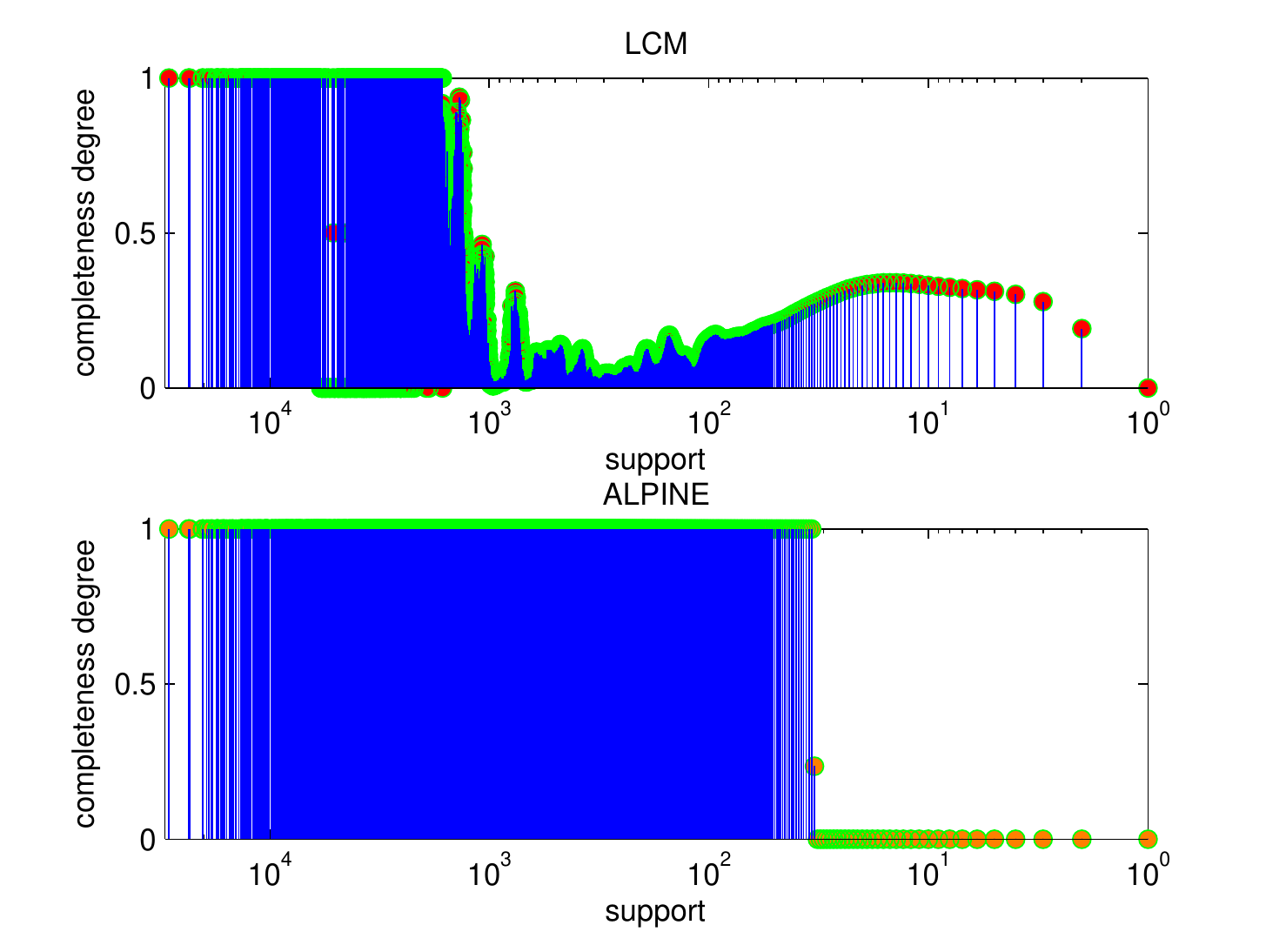}
                \caption{Hour 3}
                \label{fig:t40i10d100k3}
        \end{subfigure}
        ~ 
        \begin{subfigure}[b]{0.49\textwidth}
                \includegraphics[width=\textwidth]{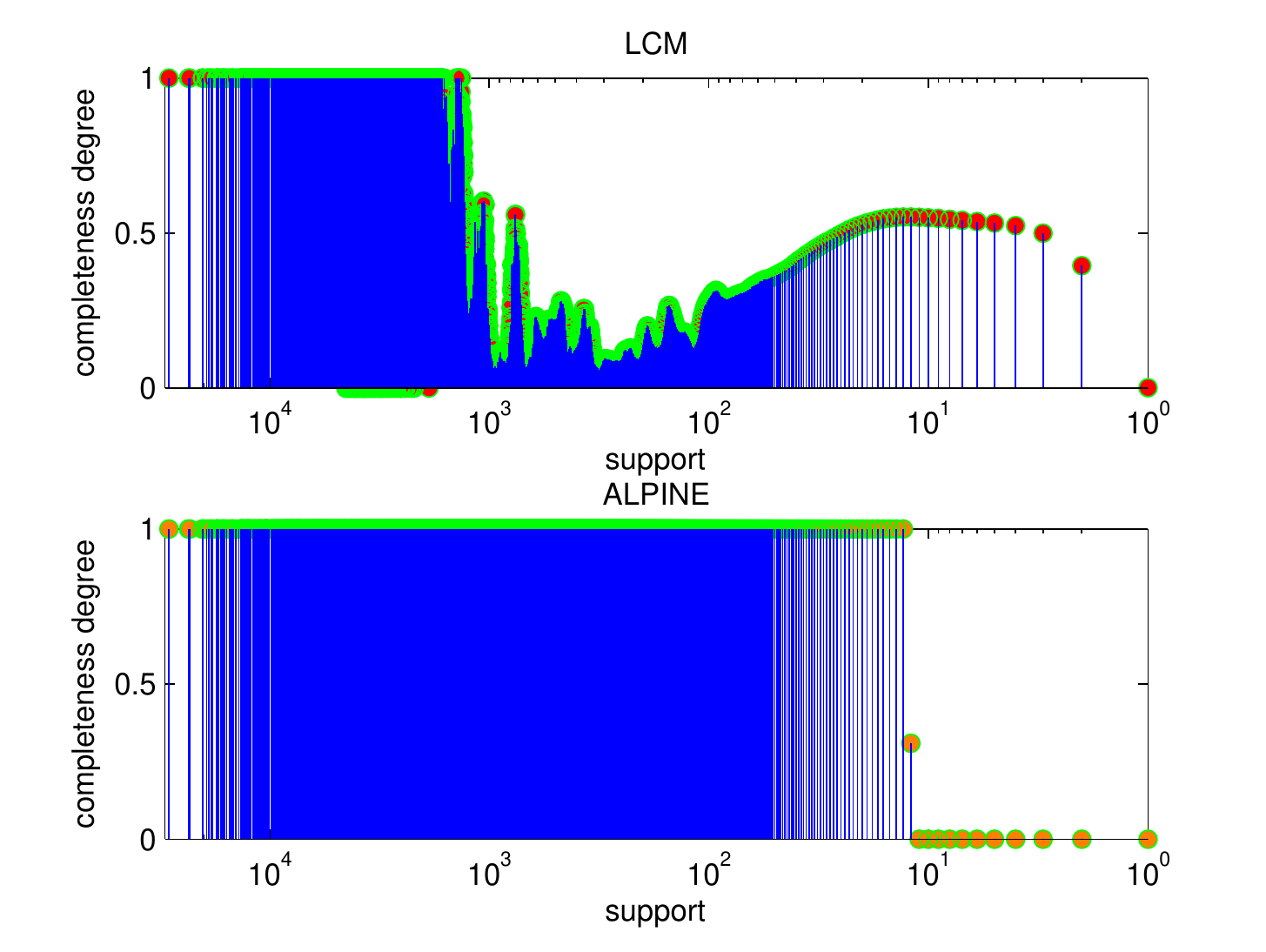}
                \caption{Hour 6}
                \label{fig:t40i10d100k6}
        \end{subfigure}
        \begin{subfigure}[b]{0.49\textwidth}
        \includegraphics[width=\textwidth]{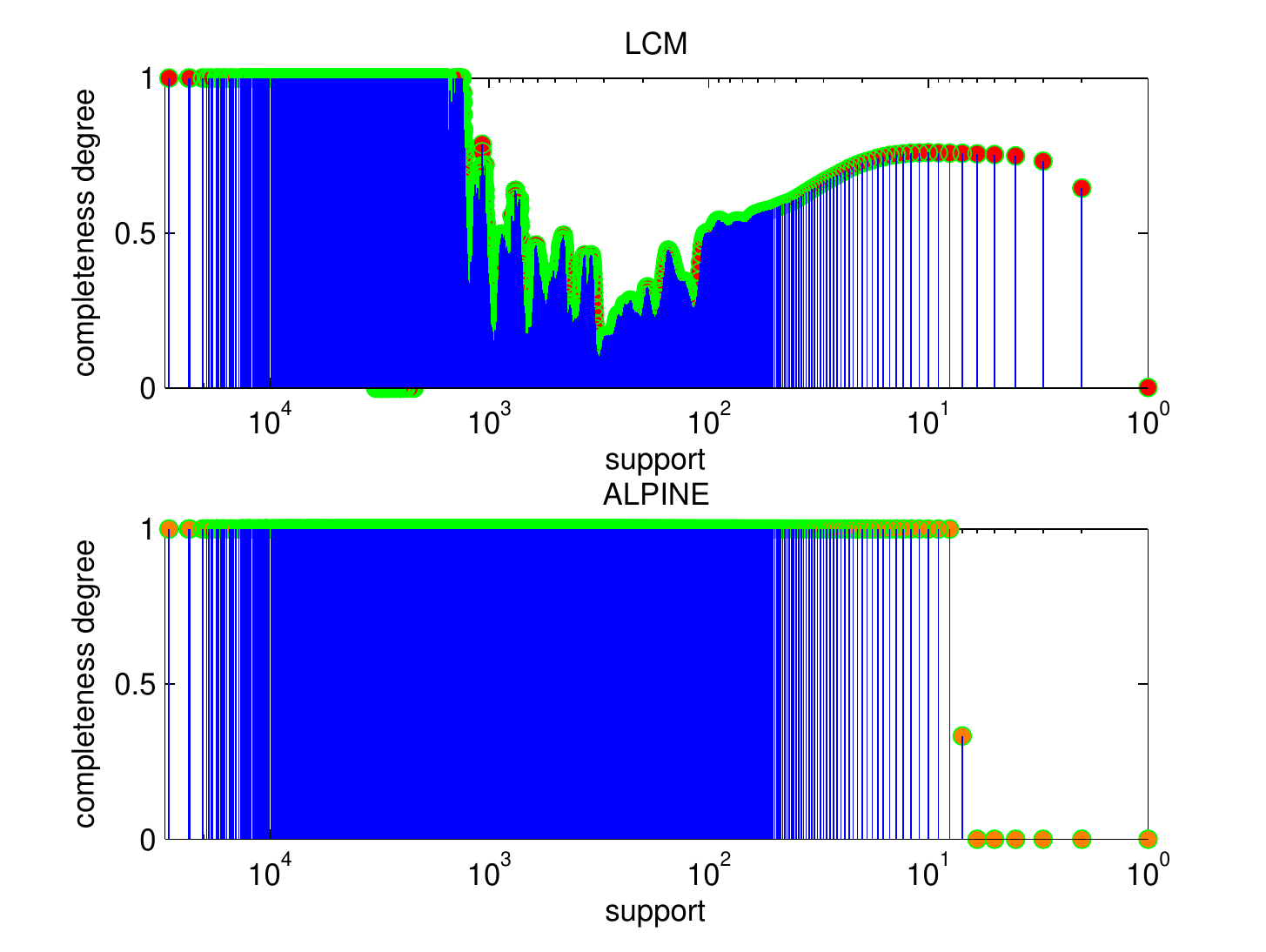}
        \caption{Hour 10}
        \label{fig:t40i10d100k10}
        \end{subfigure}
       \caption{Progress of the LCM and the ALPINE algorithm at probes on the T40I10D100K dataset. The horizontal axis is the support in decreasing order in a log scale and the vertical axis is the degree of completeness of each distinct support.}\label{fig:t40i10d100k}
\end{figure*}

The results are plotted in Figure~\ref{fig:t40i10d100k}. In each graph, the horizontal axis is the support value in a log scale, and  the vertical axis is the normalized degree of completeness. For example, in Figure~\ref{fig:t40i10d100k1}, it plots the partial answer generated by LCM and ALPINE after one hour. For LCM, in this intermediate solution, we can find almost all itemsets with different support values exist but the majority of them are incomplete. Different from LCM, in ALPINE's partial output, all larger supports to the left of the current working bin are complete, while none of the itemsets from a lower support bin have been generated. Thus, the computational overhead at each checkpoint of ALPINE is minimum.

By checking all the graphs in Figure~\ref{fig:t40i10d100k} together, we can intuitively perceive how both algorithms make progress as the computational time increases. The quality of the solution from ALPINE improves as the built support index is more and more complete. For LCM, though the completeness of a specified support value improves, in terms of the moving of minimum support, the progress is not so obvious. Imagine a dataset with even more items, the LCM algorithm might be stuck computing while ALPINE can report useful and actionable knowledge in time through \emph{checkpoints}. In the next subsection, we'll test both algorithms on a real gene expression dataset. 

\subsubsection{Experiment on gene expression dataset }
In the second set of experiments, we use LCM and ALPINE to mine all the co-regulated genes or gene groups from a real gene expression dataset from the Cancer Cell Line Encyclopedia project for the drug sensitivity analysis. The gene-centric RMA-normalized mRNA expression data consists of the expression values of 18,988 genes in 1,037 patients. To make the dataset usable for binary pattern mining algorithms, each column pertaining to the expression of a single gene is split into several binary columns. Since the data has been properly normalized, we simply adopt the equal-depth (frequency) partitioning method to discretize each gene expression  into five bins. The resulting transaction database has 94,940 items and 1,037 transactions, with a density of 20 percent. We name this dataset as the CCLE\_Expression dataset in the following paragraphs. To compress the output from this high-dimensional dataset, only \emph{closed} patterns are mined in this experiment. 

We ran both LCM and ALPINE on the gene expression dataset. The minimum support threshold of LCM is set to 80, due to the huge number of resulting closed frequent patterns from this dataset. Similar to the experimental setting in Section~\ref{section:fimi}, a series of random probes are selected in time and the minimum support reached by both algorithms are checked at every probe. The results are presented in Table~\ref{table:ccle_expression}. We can find that ALPINE can continuously make progress in terms of lowering the reached minimum support. However, the LCM algorithm is stuck in this case at the minimum support of 208. Since ALPINE always focuses on building the uncompleted bin with the highest support from the index, while LCM spreads its power to the whole support spectrum, making all bins to be completed almost at the same time. 

\begin{table*}[htb]
\caption{The $minsup$s reached at probes (in hour) by the LCM and ALPINE algorithm on the CCLE\_Expression dataset.}
\begin{tabular}{| l | m{.8cm} | m{.8cm} | m{.8cm} | m{.8cm} | m{.8cm} | m{.8cm} | m{.8cm} | m{.8cm} | m{.8cm} | m{.8cm} | m{1cm} | m{1cm} | m{1cm} |}
  \hline	
 Probe & 2 & 3 & 4 & 5 & 6 & 7 & 8 & 9 & 10 & 13 & 15 & 18 & 23 \\	
 \hline
 \hline
 LCM  & 209 & 208 & 208 & 208 & 208 & 208 & 208 & 208 & 208 & 208 & 208 & 208 & 208 \\
 \hline
 ALPINE & 136 & 125 & 122 & 120 & 119 &  118 & 117 & 116 & 115 & 114 & 113 & 112 &  111 \\
  \hline 
\end{tabular}
\label{table:ccle_expression}
\end{table*}

\begin{figure*}[t]
        \centering
        \begin{subfigure}[b]{0.33\textwidth}
                \includegraphics[width=\textwidth]{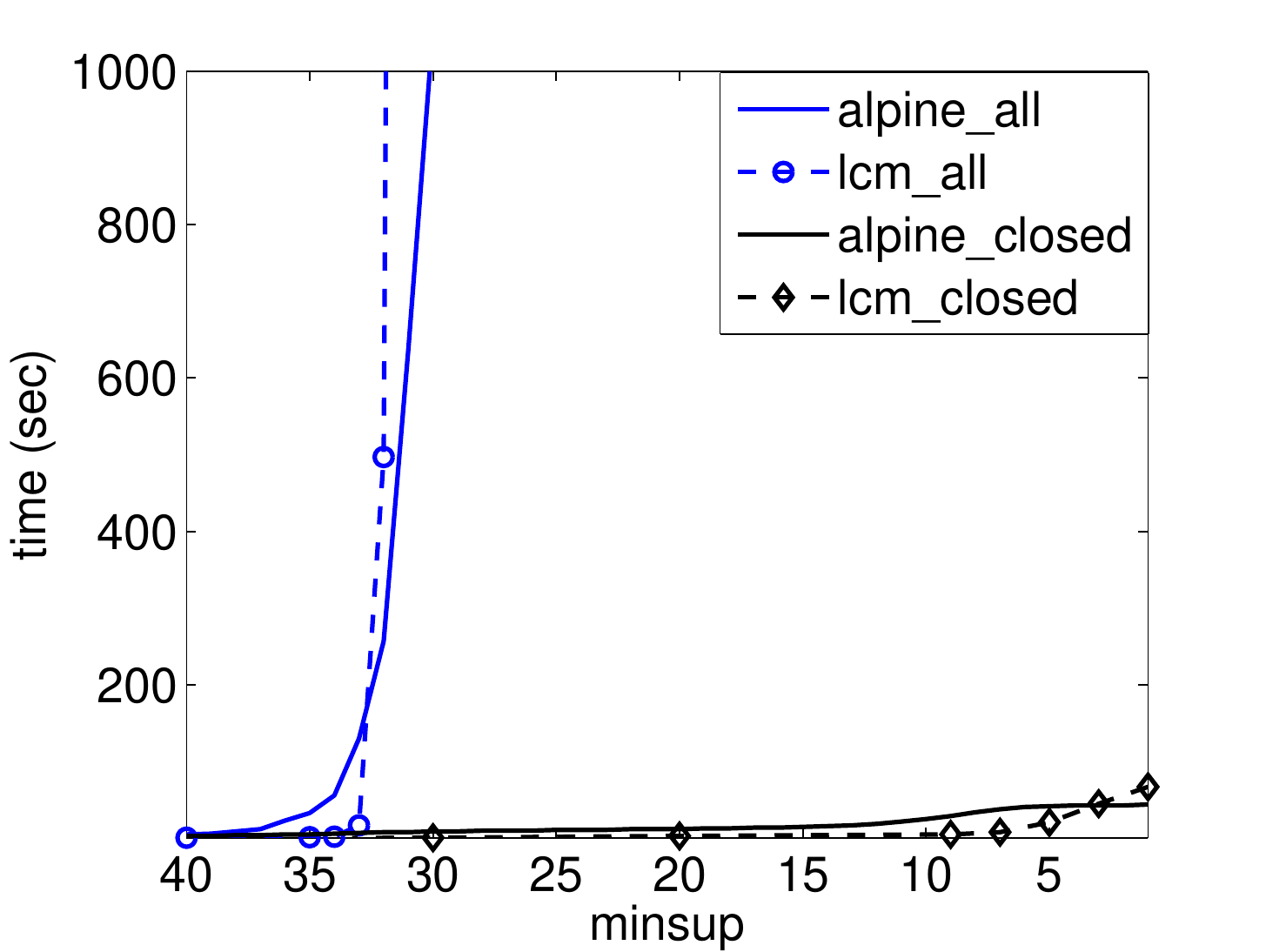}
                \caption{BMS-WebView-1}
                \label{fig:bms1}
        \end{subfigure}%
        \begin{subfigure}[b]{0.33\textwidth}
                \includegraphics[width=\textwidth]{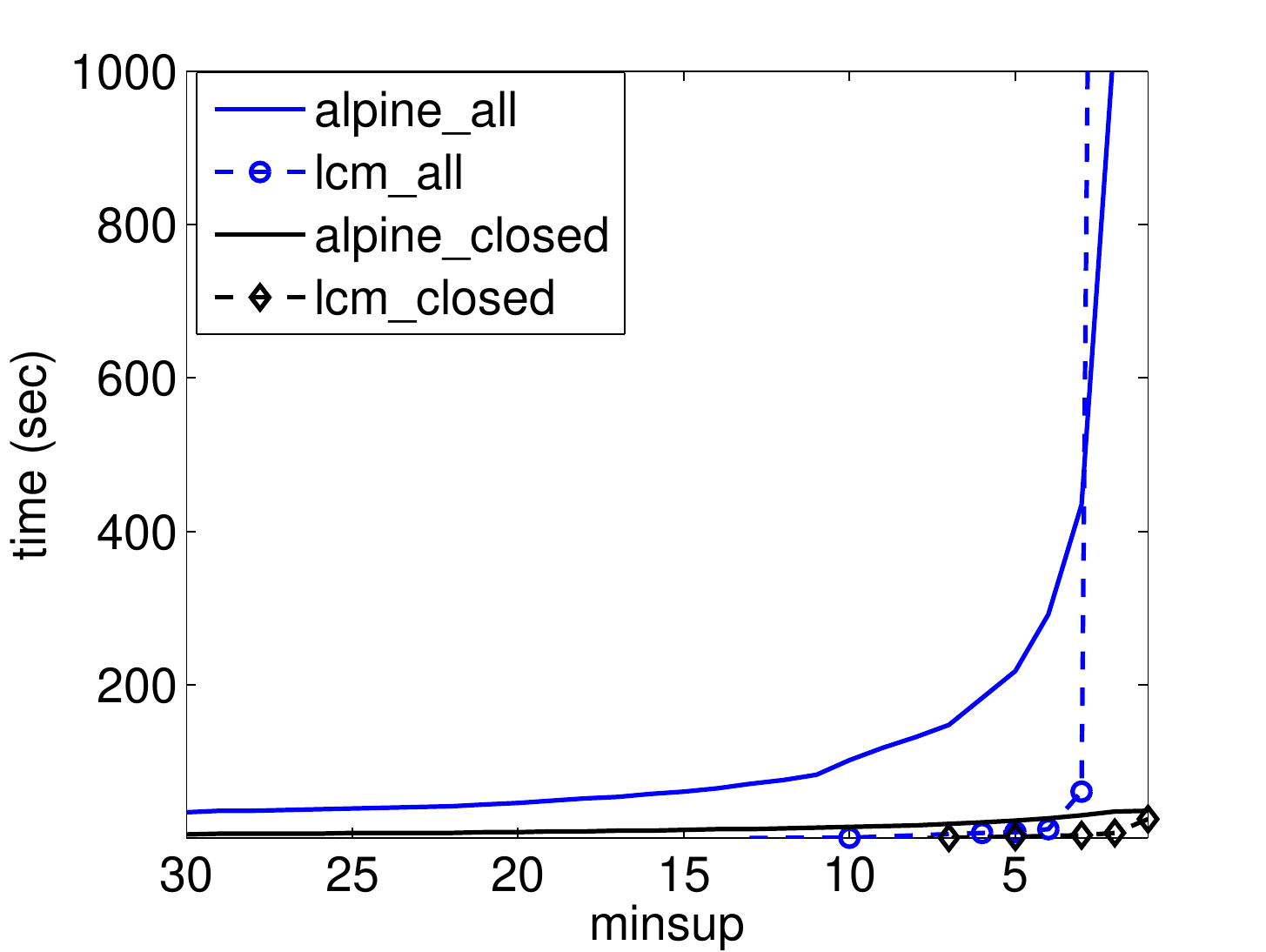}
                \caption{BMS-WebView-2}
                \label{fig:bms2}
        \end{subfigure}
        \begin{subfigure}[b]{0.33\textwidth}
                \includegraphics[width=\textwidth]{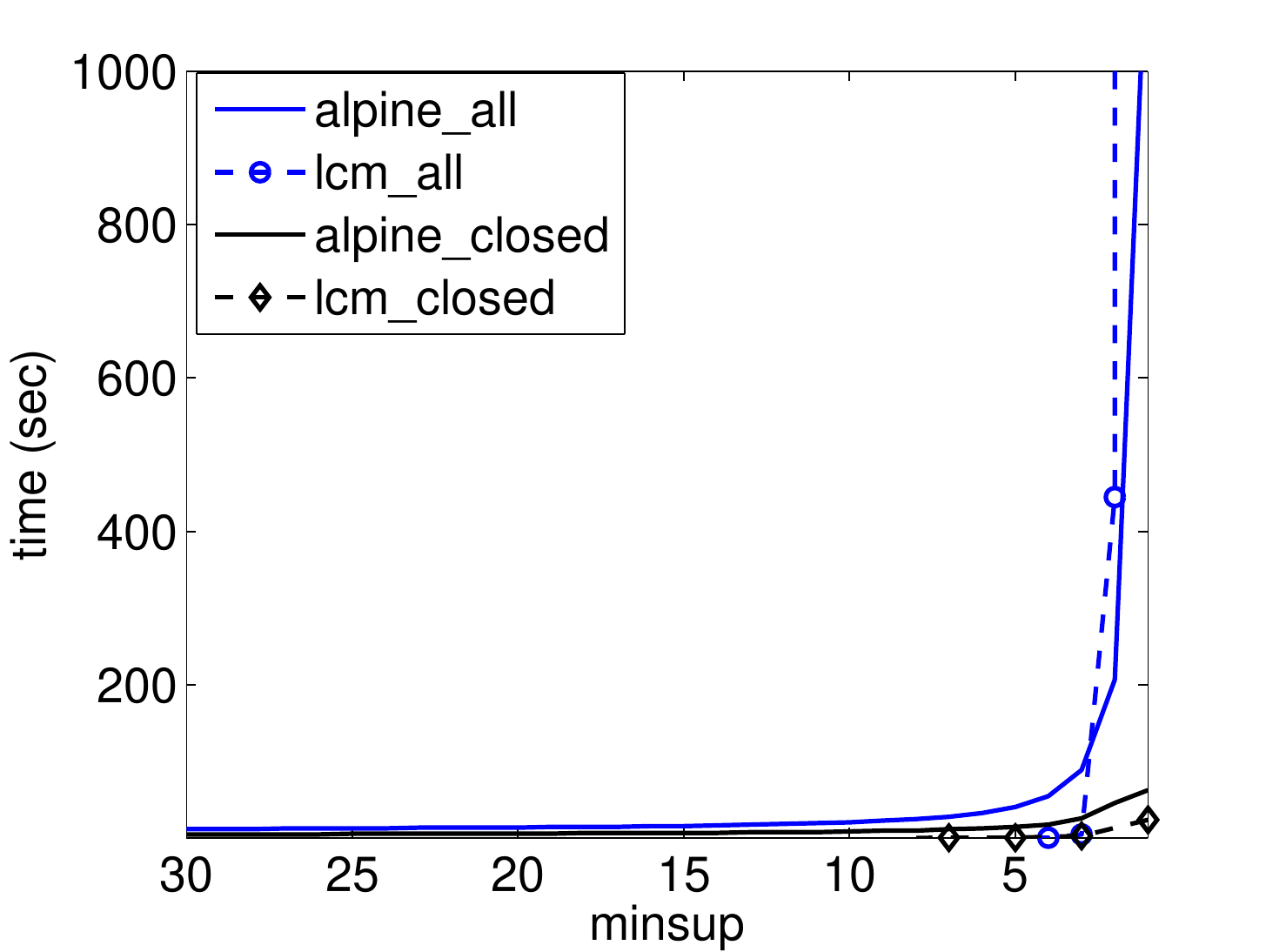}
                \caption{Retail}
                \label{fig:retail}
        \end{subfigure}
        \vspace{1mm}
        \begin{subfigure}[b]{0.33\textwidth}
                \includegraphics[width=\textwidth]{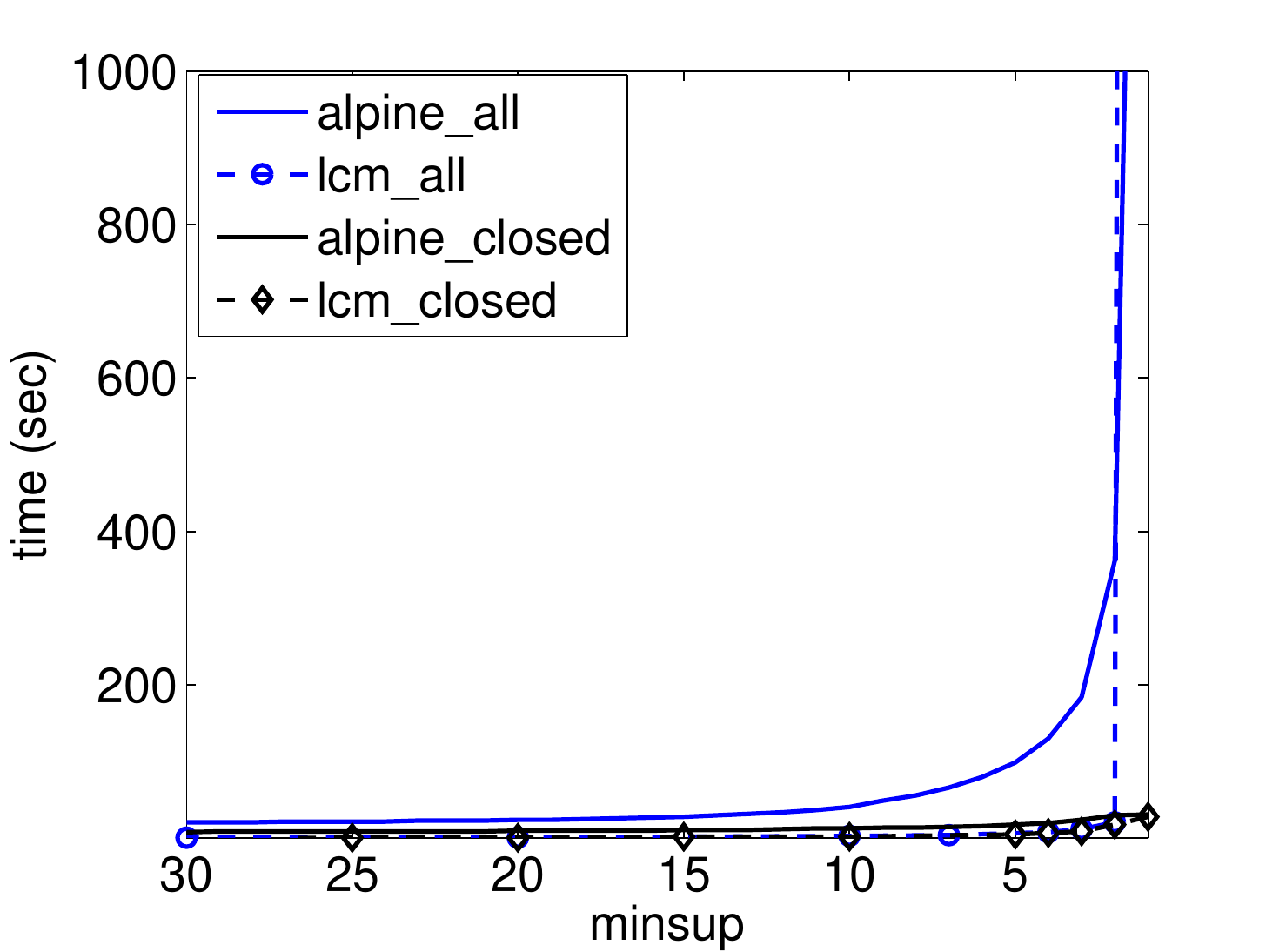}
                \caption{T10I4D100K}
                \label{fig:t10i4d100k}
        \end{subfigure}
        \begin{subfigure}[b]{0.32\textwidth}
                \includegraphics[width=\textwidth]{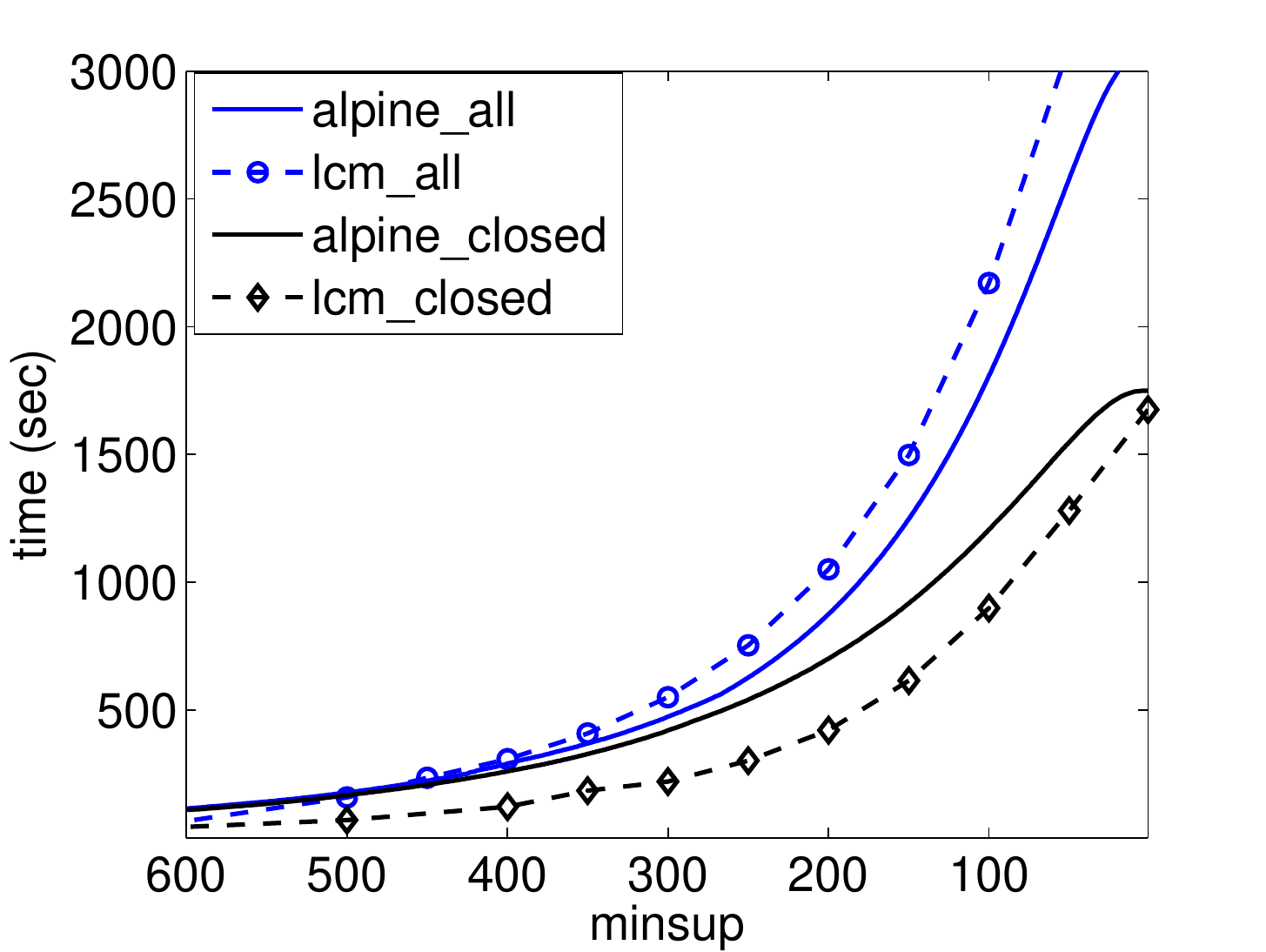}
                \caption{Chess}
                \label{fig:chess}
        \end{subfigure}
        \begin{subfigure}[b]{0.33\textwidth}
                \includegraphics[width=\textwidth]{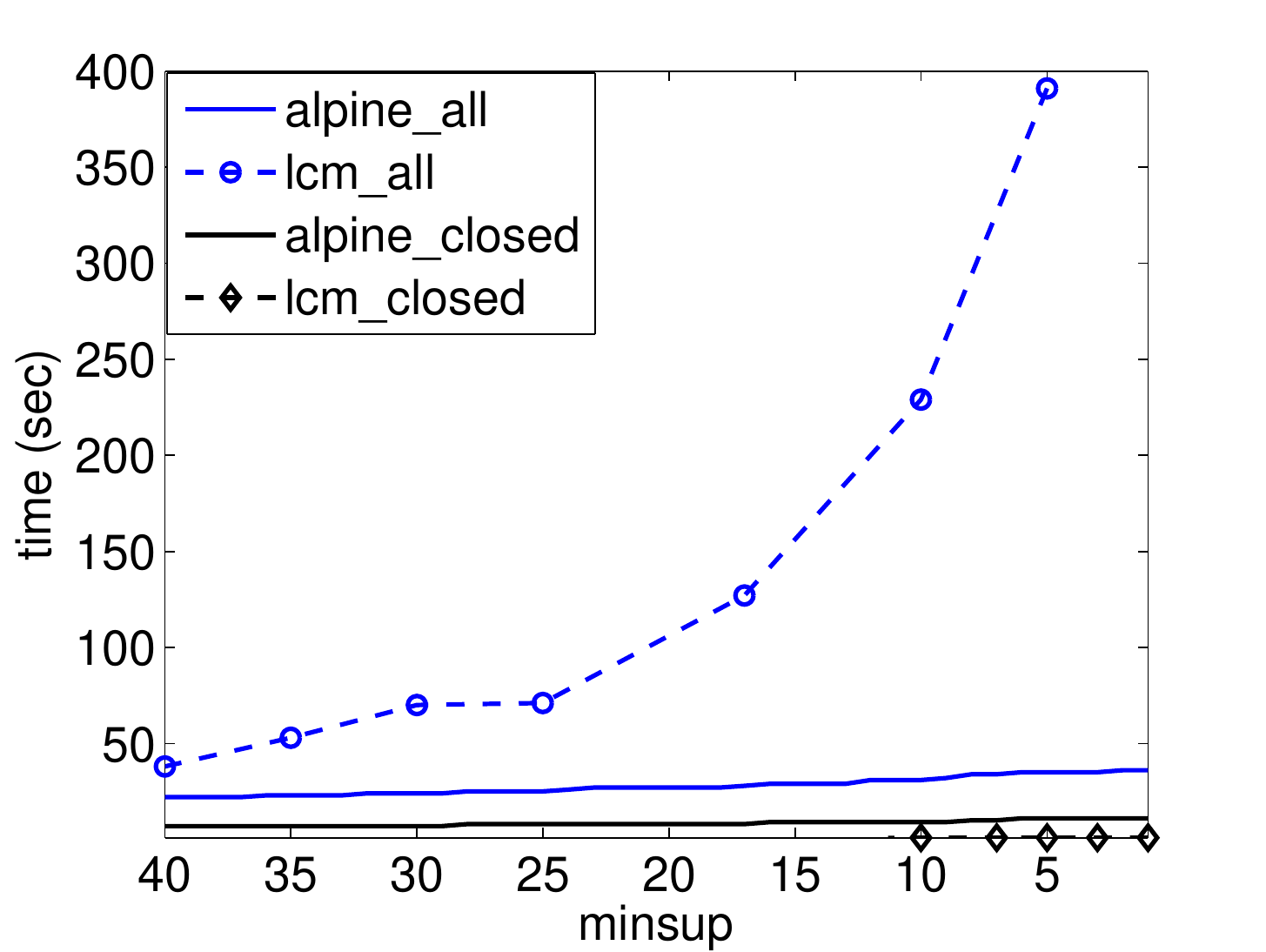}
                \caption{Mushroom}
                \label{fig:mushroom}
        \end{subfigure}            
        \vspace{1mm}
       \caption{Computational overhead of ALPINE}\label{fig:fim}
\end{figure*}

In practical scenarios like this one, it is generally reasonable to have some time period for most data analysis operations. Though ALPINE might also not be able to finish the whole mining task within the time period, but it is always able to offer a partial solution based on its last checkpoint. Furthermore, the partial solution offered by ALPINE is \emph{complete} in itself and has the definite guarantee with regard to a higher minimum support. Thus, these complete sets of co-regulated genes or gene groups with a higher minimum support returned early by ALPINE can be used to predict the drug response even though the mining process continues. The high support implies high coverage, which might lead to more widely applicable associations in this case. Besides, as the computational time increases, the built support index is more and more complete and ALPINE continues to offer those lower support patterns. 

\subsubsection{Computational overhead of ALPINE}
\label{section:overhead}

Many frequent pattern discovery algorithms have been developed in literature and it is not our intention to develop yet another efficient algorithm for finding these patterns. Instead, our aim here is to show the usefulness of anytime data mining. To complete the picture, we also conducted experiments to evaluate the performance of ALPINE in mining frequent/closed itemsets in comparison with LCM. 

In this set of experiments, we select BMS-WebView-1, BMS-WebView-2, Retail, T10I4D100K, Chess and Mushroom datasets from the FIMI repository. ALPINE started without any parameter, while LCM was initialized with some minimum support value from ALPINE's checkpoints for the comparison purpose. The results are displayed in Figure~\ref{fig:fim}. In each graph, the horizontal axis is the absolute minimum support value, and the vertical axis is the runtime. Note that for every transaction database, ALPINE executes once to mine all frequent or closed itemsets, while LCM runs multiple times for the set of different initial minimum support values. The curves for alpine\_all and alpine\_closed in the plots are continuous in the sense that the runtime is known for each distinct $minsup$ value, indicated by solid lines. In contrast, the lcm\_all and lcm\_closed are plotted in dashed lines for only the results at markers were tested.

Overall, for all instances and minimum support values, the ALPINE algorithm is comparable to the LCM algorithm, which can be verified from the graphs, though the extra support index information needs to be maintained. The results validate the effectiveness of the itemset closure operator and the compact itemset interval representation. Generally, the runtime grows much slower for closed itemset mining than that of all frequent itemset mining. For closed itemset mining, ALPINE is slightly slower than LCM, however, the trend and the order of magnitude of the runtime of both algorithms are similar. For all frequent itemset mining, ALPINE catch up with or even compete with LCM as they get to lower and lower minimum support. The reason is ALPINE processes and outputs groups of itemsets compactly in itemset intervals  instead of enumerating each individual itemset in an interval.

\begin{figure*}[t]
        \centering
        \begin{subfigure}[b]{0.33\textwidth}
                \includegraphics[width=\textwidth]{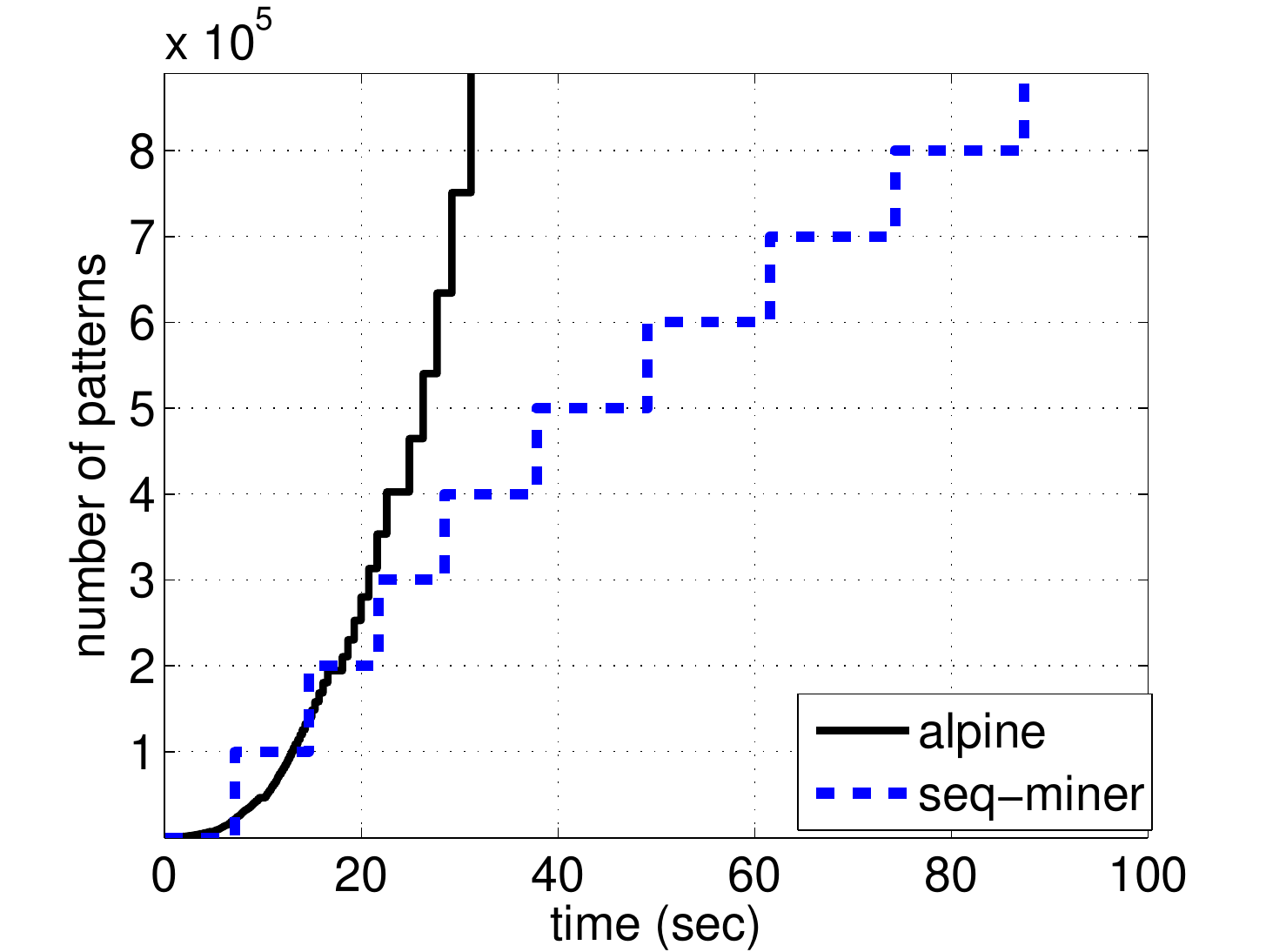}
                \caption{BMS-WebView-2}
                \label{fig:topk-bms2}
        \end{subfigure}%
        \begin{subfigure}[b]{0.33\textwidth}
                \includegraphics[width=\textwidth]{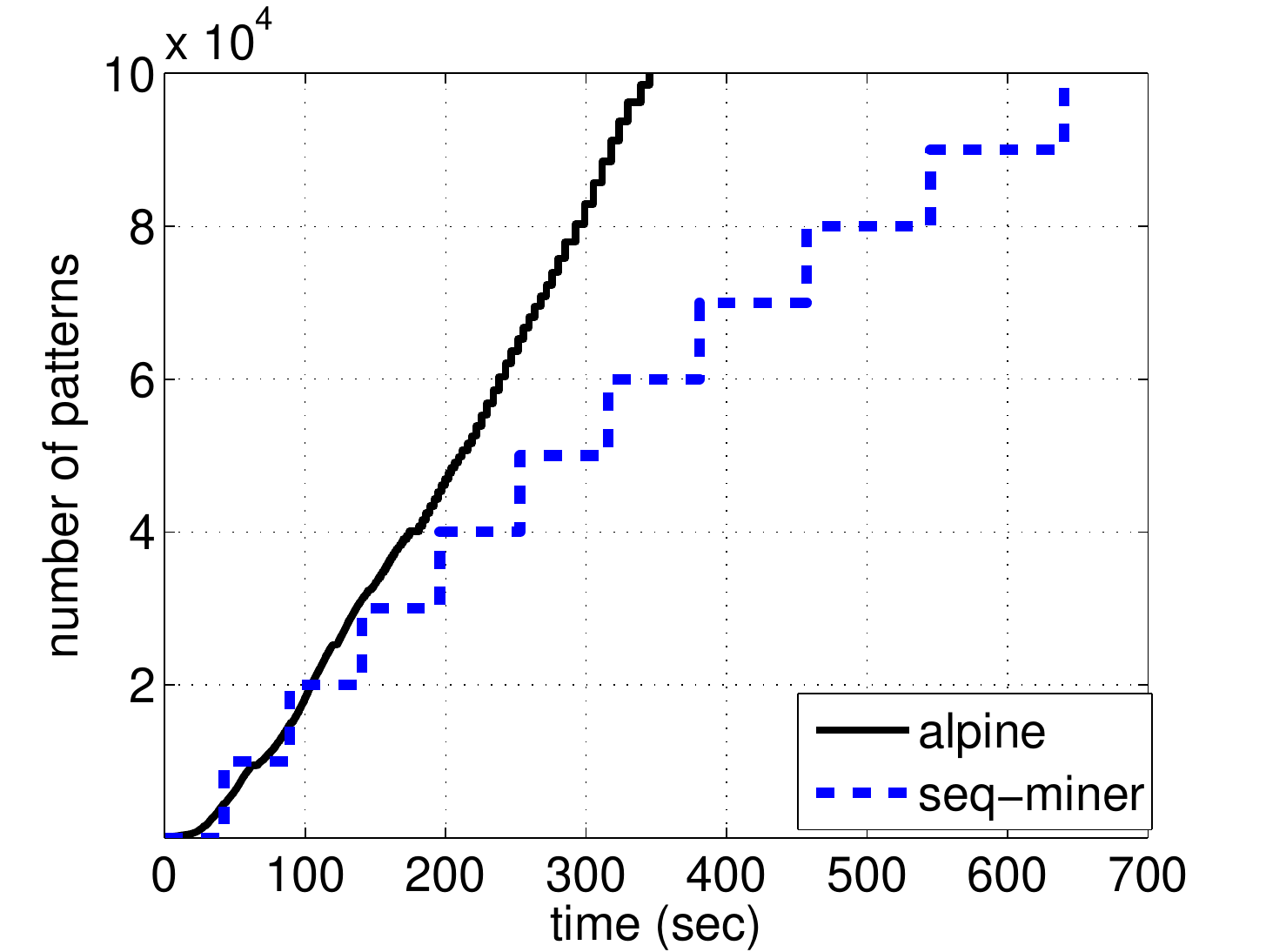}
                \caption{T40I10D100K}
                \label{fig:topk-t40i10d100k}
        \end{subfigure}%
        \begin{subfigure}[b]{0.33\textwidth}
                \includegraphics[width=\textwidth]{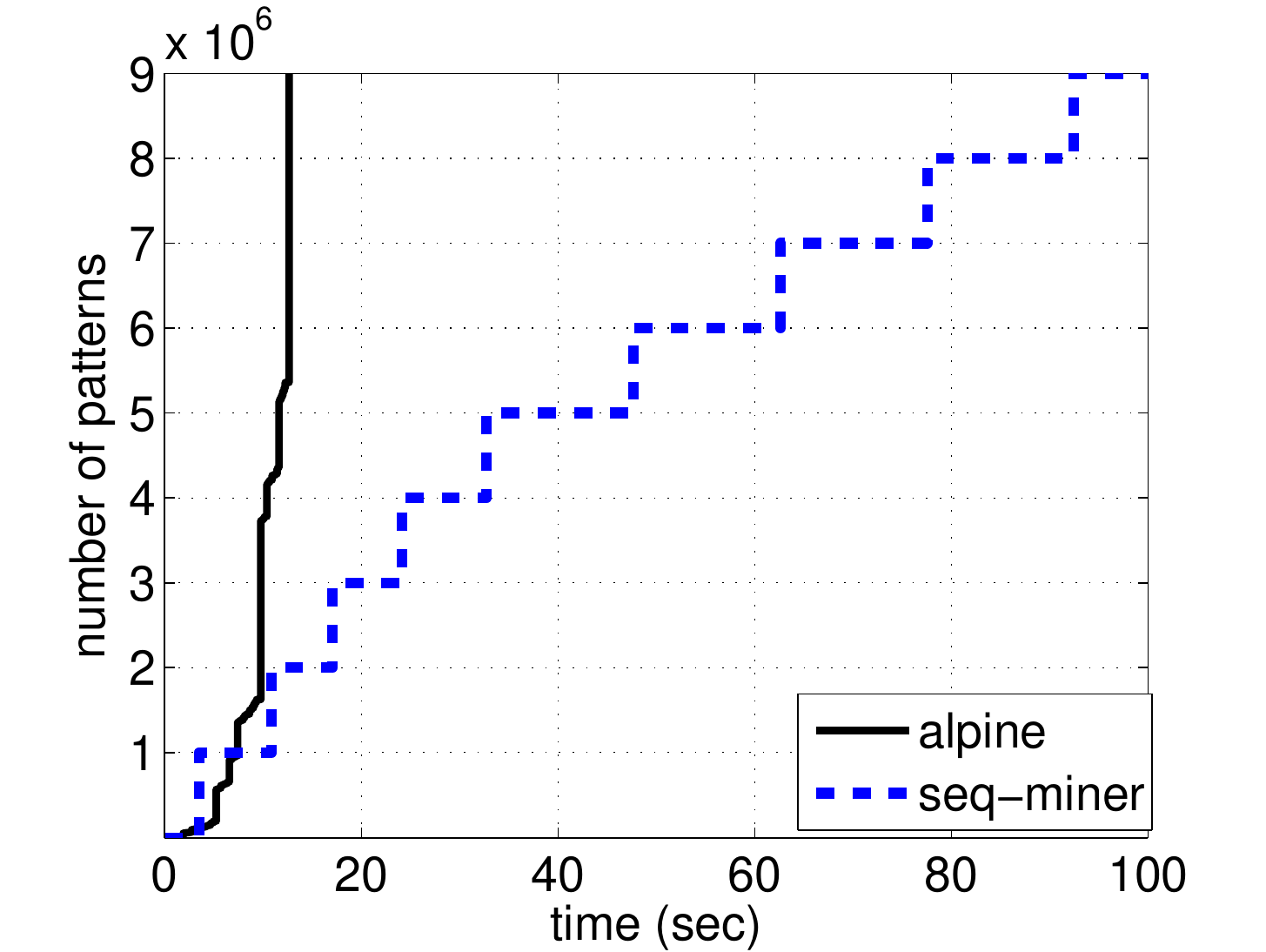}
                \caption{Mushroom}
                \label{fig:topk-mushroom}
        \end{subfigure}%
       \caption{Comparison with sequential top-$k$ mining algorithm}\label{fig:topk}
\end{figure*}

For sparse datasets like Figure~\ref{fig:bms1} - Figure~\ref{fig:t10i4d100k}, the graphs show similar trends: the curves of LCM and ALPINE are close to each other and ALPINE is slightly slower than LCM at the beginning. The difference between them might further increase in the middle of the curves as they are mining all frequent itemsets, for the overhead in generating and maintaining a large number of itemset intervals below the current $minsup$ might dominate the acceleration of the itemset closure operator and itemset interval compression. However, as the $minsup$ gets lower and lower, the previously built partial index for lower minimum supports saves the cost for later stages and the low support itemsets might be more compactly grouped in itemset intervals. That's why we can see ALPINE might compete with LCM at some lower minimum support value for all frequent itemset mining on these datasets. For dense datasets in Figure~\ref{fig:chess} and Figure~\ref{fig:mushroom}, the compression ratio of itemset intervals is even higher, so ALPINE becomes faster than LCM for mining all frequent itemsets. This advantage will become more obvious as we get to lower and lower minimum support, as indicated in Figure~\ref{fig:mushroom}.

\subsection{Comparison with sequential top-$k$ mining}
In this subsection, we also compared the performance of ALPINE with Seq-Miner. Seq-Miner mines the top-$k$ frequent patterns sequentially and outputs every top $n_c$ (a user defined chunk size) patterns. Seq-Miner shares some flavor of the contract-type anytime algorithm and it can also provide definite results at each chunk, i.e, all itemsets with support above or equal to the support of the last one in the chunk. The advantage is similar to the one provided by ALPINE, checkpoints can now be provided just like ALPINE. 

From the set of datasets used in Section~\ref{section:fim}, we selected two sparse datasets and one dense dataset as representatives for this experiment, namely, BMS-WebView-2, T40I10D100K and Mushroom. For Seq-Miner, the chunk size $n_c$ of these datasets are selected to be $10^5$, $10^4$ and $10^6$, respectively, according to the density and output number of  frequent patterns of each dataset. The number of patterns generated at each checkpoint and the time to reach that checkpoint is plotted in Figure~\ref{fig:topk} for both algorithms. In this figure, the horizontal axis is the running time since the algorithm starts and the vertical axis is the number of generated patterns. Between any two of these checkpoints, the intermediate status is unchecked, that's why we got these staircase-shaped curves in Figure~\ref{fig:topk}. 

From these graphs, it's easy to verify the following facts: 1) ALPINE produces far more number of checkpoints than Seq-Miner given the same execution time; 2) the step size (time between two successive checkpoints) of Seq-Miner is much longer than that of ALPINE, and it grows as the running time increases. The reason is a new and larger FP-tree has to be rebuilt from scratch whenever a given top-$k$ is changed in the Seq-Miner. Every time a new call to the mining algorithm is made with the smaller value of minimum support discovered in the VirtualGrowth. Thus, the FP-tree is built many times and the most frequent itemsets are generated again and again. Different from Seq-Miner, ALPINE monotonically explores itemset intervals with descending values of support and mines continuously from checkpoint to checkpoint without any redundancy (never starting from scratch). Not surprisingly, the iterative process of Seq-Miner incurs substantial time penalty as compared to that of ALPINE. 

The step size of Seq-Miner is related with the parameter - chunk size $n_c$, and we can reduce the step size by reducing its chunk size. In that case, it will result in more iterations in this iterative process and more repeated work in total. In general, the number of iterations and the total runtime of Seq-Miner might grow dramatically as we generate more and more frequent patterns. This is consistent with the trend displayed in the graphs of Figure~\ref{fig:topk} that the runtime difference of Seq-Miner and ALPINE grows with the increasing of the number of generated patterns. Thus, the superiority of ALPINE increases with the number of iterations of frequent pattern mining of Seq-Miner. Given the same time, ALPINE can always generate more frequent patterns than Seq-Miner. In other words, using ALPINE, users can obtain the complete set of itemsets above a lower $minsup$ in the equivalent execution time in comparison with using Seq-Miner. ALPINE turns out to be even more efficient than the contract-type like algorithm, though it is interruptible at any time.

\section{Related Work}
\textbf{Frequent itemset mining}: A lot of algorithms have been proposed to mine itemsets in the past decade~\cite{agrawal94:apriori, deng12:nlists, deng15:prepost+, han04:fpgrowth, pasquier99:closed, uno03:lcm, zaki00:eclat}, the key is how to efficiently reduce the search space. Apriori-like methods utilize the anti-monotone property to prune candidates~\cite{agrawal94:apriori, zaki00:eclat}, FP-growth family employs some highly condensed data structure, such as FP-tree~\cite{han04:fpgrowth} or PPC-tree~\cite{deng12:nlists}, to confine the search space, while PrePost+~\cite{deng15:prepost+} introduces the children-–parent equivalence pruning strategy. However, the pruning might be incomplete. Thus, the closure operator $I(\mathcal{T}(\cdot))$ is incorporated in other algorithms~\cite{pasquier99:closed, uno03:lcm, uno05:lcmv3} as we do. In\cite{pasquier99:closed}, duplicated closed itemset may be generated. The most similar work to ours is the LCM algorithm~\cite{uno03:lcm, uno05:lcmv3}, which also transverses a tree composed of closed itemsets. However, LCM requires to set the $minsup$ threshold and no completeness guarantees is given for its intermediate partial solutions. 

\textbf{Top-$k$ mining}: In \lq\lq concept mining\rq\rq, the top-$k$ mining can gradually raise the $minsup$ to mine \lq\lq the $k$-most interesting patterns\rq\rq~without specifying a $minsup$ threshold in advance to increase the usability of a data mining algorithm. Shen \textit{et al.}~\cite{shen98:nlargest} first introduced the top-$k$ mining problem to generate an appropriate number of most interesting itemsets. The Itemset-Loop/Itemset-iLoop algorithm~\cite{fu00:miningn-most} based on the Apriori approach~\cite{agrawal94:apriori} and the TFP algorithm~\cite{wang05:tfp} extending the FP-growth method~\cite{han04:fpgrowth} are developed to mine the $k$-most interesting patterns thereafter. Generally, these algorithms follow the same process: Initially, the $minsup$ threshold is set to 0 to ensure no pattern will be missing, then the $minsup$ is gradually raised by the algorithm to prune the search space until top-$k$ patterns are found. Though these algorithms don't need the parameter $minsup$, but the threshold $k$ is still necessary. When $k$ is too large, mining takes an unacceptably long time; on the contrary, when $k$ is too small, it will miss a lot of potential interesting patterns. The problem of setting up the value of $minsup$ is now replaced with setting the value of $k$. Thus, Hirate \emph{et. al.}~\cite{hirate04:tf2p-growth} propose the TF$^2$P-growth algorithm to mine the top-$k$ pattens sequentially without any thresholds. 

TF$^2$P-growth outputs every top $n_c$ patterns, where $n_c$ is some user-defined chunk size. For instance, $n_c = 1000$, it sequentially returns exactly the top 1000, 2000, 3000 patterns \textit{etc}. For $n_c$ is a user specified number, it might not return all itemsets having the same support as the last one. Minh \emph{et al.} \cite{minh06:seq-miner} overcame this shortcoming and proposed an improved algorithm, the Seq-Miner. These methods have a flavor of the contract-type anytime algorithms~\cite{zilberstein96:anytime}, but they can not be interrupted before the termination of every top $n_c$ patterns. In contrast, ALPINE monotonically explores itemsets with descending values of support and mines continuously from checkpoint to checkpoint, which guarantees the quality of the partial results is checked at any time.

\textbf{Pattern sampling}: Zhang \emph{et al.}~\cite{zhang02:anytime} used sampling and incremental mining to support multiple-user inquiries at any time. Boley \emph{et al.}~\cite{boley10:formalConcept, boley11:direct} proposed to use Metropolis-Hastings sampling for the construction of data mining systems that do not require any user-specified threshold, \emph{i.e.}, minimum support or confidence. However, all the algorithms generate approximate results and the completeness cannot be guaranteed. 

\section{Conclusion}


In this work, we defined the anytime itemset mining problem and proposed the ALPINE algorithm. ALPINE proceeds in the defined anytime mining manner and can be interrupted at any time but offer intermediate meaningful and complete results with definite guarantees. ALPINE is, to our knowledge, the first interruptible anytime algorithm to mine frequent itemsets and closed frequent itemsets. It guarantees that all itemsets with support exceeding the current checkpoint's support have been found before it proceeds further. This ANYTIME feature is the most important contribution of ALPINE, which is also fast but not necessarily the fastest algorithm around. Another critical advantage of ALPINE is that it do not require setting the minimum support apriori, but can be adjusted automatically as the mining process continues.


\begin{thebibliography}{20}

\bibitem{agrawal94:apriori}
R.~Agrawal and R.~Srikant.
\newblock Fast algorithms for mining association rules.
\newblock In {\em Proceedings of 20th Int. Conf. on VLDB}, pages
  487--499, 1994.

\bibitem{arai07:anytimetopk}
B.~Arai, G.~Das, D.~Gunopulos, and N.~Koudas.
\newblock Anytime measures for top-k algorithms.
\newblock In {\em Proceedings of the 33rd International Conference on Very
  Large Data Bases}, VLDB '07, pages 914--925, 2007.

\bibitem{boley10:formalConcept}
M.~Boley and T.~G{\"a}rtner and H.~Grosskreutz. 
\newblock Formal Concept Sampling for Counting and Threshold-Free Local Pattern Mining.
\newblock In {\em SIAM Int. Conf. on Data Mining}, pages 177--188, 2010.

\bibitem{boley11:direct}
M.~Boley and C.~Lucchese and D.~Paurat and T.~G{\"a}rtner.
\newblock Direct local pattern sampling by efficient two-step random procedures.
\newblock In {\em Int. Conf. on Knowledge Discovery and Data Mining}, pages 582--590, 2011.

\bibitem{borgelt12:fism}
C.~Borgelt.
\newblock Frequent item set mining.
\newblock In {\em Wiley Interdisciplinary Reviews: Data Mining and Knowledge
  Discovery}, 2(6):437--456, 2012.

\bibitem{dass05:bdfs}
R.~Dass and A.~Mahanti.
\newblock Fast frequent pattern mining in real-time.
\newblock In {\em Proceedings of the Eleventh International Conference on
  Management of Data}, pages 156--167, 2005.

\bibitem{deng12:nlists}
Z.~Deng, Z.~Wang, and J.~Jiang.
\newblock A new algorithm for fast mining frequent itemsets using n-lists.
\newblock In {\em Sci. China, Inf. Sci.}, 55(9):2008--2030, 2012.

\bibitem{deng15:prepost+}
Z.~H. Deng and S.~L. Lv.
\newblock Prepost+: An efficient n-lists-based algorithm for mining frequent
  itemsets via children-parent equivalence prunings.
\newblock In {\em Expert Syst. Appl.}, 42(13):5424--5432, 2015.

\bibitem{fu00:miningn-most}
A.~W. chee Fu, R.~W. wai Kwong, F.~Renfrew, W.~wai Kwong, and J.~Tang.
\newblock Mining n-most interesting itemsets, 2000.

\bibitem{han04:fpgrowth}
J.~Han, J.~Pei, Y.~Yin, and R.~Mao.
\newblock Mining frequent patterns without candidate generation: A
  frequent-pattern tree approach.
\newblock {\em Data Min. Knowl. Discov.}, 8(1):53--87, Jan. 2004.

\bibitem{hirate04:tf2p-growth}
Y.~Hirate, E.~Iwahashi, and H.~Yamana.
\newblock Tf$^2$p-growth: An efficient algorithm for mining frequent patterns
  without any thresholds.
\newblock In {\em Proceedings of IEEE International Conference on Data Mining},
  2004.

\bibitem{kranen09:anytime}
P.~Kranen and T.~Seidl.
\newblock Harnessing the strengths of anytime algorithms for constant data  streams.
\newblock {\em Data Min. Know. Dis.}, 19(2):245--260, 2009.

\bibitem{minh06:seq-miner}
Q.~T. Minh, S.~Oyanagi, and K.~Yamazaki.
\newblock Mining the k-most interesting frequent patterns sequentially.
\newblock In {\em Proceedings of the 7th International Conference on IDEAL},
  pages 620--628, 2006.

\bibitem{pasquier99:closed}
N.~Pasquier, Y.~Bastide, R.~Taouil, and L.~Lakhal.
\newblock Discovering frequent closed itemsets for association rules.
\newblock In {\em Proceedings of 7th International Conference on Database
  Theory (ICDT '99)}, pages 398--416, 1999.

\bibitem{rothbard56:law}
M.~N. Rothbard.
\newblock Toward a reconstruction of utility and welfare economics.
\newblock 1956.

\bibitem{shen98:nlargest}
L.~Shen, H.~Shen, P.~Pritchard, and R.~Topor.
\newblock Finding the n largest itemsets.
\newblock In {\em Int'l Conf. on Data Mining}, pages 211--222,
  1998.

\bibitem{uno03:lcm}
T.~Uno, T.~Asai, Y.~Uchida, and H.~Arimura.
\newblock Lcm: An efficient algorithm for enumerating frequent closed item
  sets.
\newblock In {\em Proceedings of Workshop on Frequent itemset Mining
  Implementations (FIMI’03)}, 2003.

\bibitem{uno05:lcmv3}
T.~Uno, M.~Kiyomi, and H.~Arimura.
\newblock Lcm ver.3: Collaboration of array, bitmap and prefix tree for
  frequent itemset mining.
\newblock In {\em the 1st Int. Workshop on Open Source
  Data Mining: Frequent Pattern Mining Implementation}, pages 77--86, 2005.

\bibitem{wang05:tfp}
J.~Wang, J.~Han, Y.~Lu, and P.~Tzvetkov.
\newblock {TFP:} an efficient algorithm for mining top-k frequent closed
  itemsets.
\newblock {\em {IEEE} Trans. Knowl. Data Eng.}, 17(5):652--664, 2005.

\bibitem{zaki00:eclat}
M.~J. Zaki.
\newblock Scalable algorithms for association mining.
\newblock {\em IEEE Trans. on Knowl. and Data Eng.}, 12(3):372--390, May 2000.

\bibitem{zaki98:theoretical}
M.~J. Zaki and M.~Ogihara.
\newblock Theoretical foundations of association rules.
\newblock In {\em In 3rd ACM SIGMOD Workshop on Research Issues in Data Mining
  and Knowledge Discovery}, 1998.

\bibitem{zhang02:anytime}
S.~Zhang and C.~Zhang. 
\newblock Anytime mining for multiuser applications.
\newblock In {\em IEEE Trans. Systems, Man, and Cybernetics}, 32(4):515--521, 2002.

\bibitem{zilberstein96:anytime}
S.~Zilberstein.
\newblock Using anytime algorithms in intelligent systems.
\newblock {\em AI Magazine}, 17(3):73--83, 1996.

\end{thebibliography}
\end{document}